\documentclass[usenatbib]{mnras}
\usepackage{xcolor}
\usepackage{hyperref}
\usepackage{enumerate}
\usepackage [caption=false]{subfig}
\usepackage{natbib}
\usepackage{url}
\usepackage{threeparttable}
\usepackage{graphicx}	
\usepackage{amsmath}	
\usepackage{amssymb}

\usepackage{mathrsfs}
\newcommand\xmm{XMM-{\it Newton} }

\defcitealias{Gallo_2021}{G21}

\title[The X-ray occultation event in NGC 6814]{What can be learnt from a highly informative X-ray occultation event in NGC 6814? A marvellous absorber}

\author[Jia-Lai Kang et al.]{
Jia-Lai Kang$^{1,2}$\thanks{ericofk@mail.ustc.edu.cn},
Jun-Xian Wang$^{1,2}$\thanks{jxw@ustc.edu.cn},
Shu-Qi Fu$^{1,2}$
\\
$^{1}$CAS Key Laboratory for Research in Galaxies and Cosmology, Department of Astronomy, University of Science and Technology of China, \\
  Hefei, Anhui 230026, China\\
$^{2}$School of Astronomy and Space Science, University of Science and Technology of China, Hefei 230026, China\
}

\pubyear{2023}

\begin{document}
\label{firstpage}
\pagerange{\pageref{firstpage}--\pageref{lastpage}}
\maketitle
\begin{abstract}
A unique X-ray occultation event in NGC 6814 during an \xmm observation in 2016 has been reported, providing useful information of the absorber and the corona. 
We revisit the event with the aid of the hardness ratio (HR) -- count rate (CR) plot and comparison with two other absorption-free XMM exposures in 2009 and 2021. 
NGC 6814 exhibits a clear ``softer-when-brighter" variation pattern during the exposures, but the 2016 exposure significantly deviates from the other two in the HR -- CR plot. 
While spectral fitting does yield transient Compton-thin absorption corresponding to the eclipse event in 2016, rather than easing the tension between exposures in the HR -- CR plot, correcting the transient Compton-thin absorption results in new and severe deviation within the 2016 exposure.
We show that the eclipsing absorber shall be clumpy (instead of a single Compton-thin cloud), with an inner denser region composed of both Compton-thin and Compton-thick clouds responsible for the previously identified occultation event, and an outer sparser region with Compton-thin clouds which eclipses the whole 2016 exposure. With this model, all the tension in the HR -- CR plots could be naturally erased, with the observed spectral variability during the 2016 exposure dominated by the variation of absorption.
Furthermore, the two warm absorbers (with different ionization and column densities but similar outflowing velocities)  detected in the 2016 exposure shall also associate with the transient absorber, likely due to ablated or tidal stretched/disrupted fragments. This work highlights the unique usefulness of the HR -- CR plot while analysing rare occultation events. 
\end{abstract}


\section{Introduction} \label{sec:intro}
\par It is widely believed the X-ray emission of active galactic nuclei is produced in a compact region near the supermassive black hole, the so called corona \citep{Haardt_1991, Haardt_1993}. Such emission can be attenuated by intervening materials, rendering altered fluxes and spectral shapes. If an obscuring cloud moves across our line of sight, an occultation/eclipse event occurs, producing unique features in the observed light curves and time-resolved spectra. Such occultation events are of particular concern, as they can provide a wealth of information about both the eclipsing absorber and the X-ray corona. To date, variations of X-ray absorption have been observed in a number of sources \citep[e.g.,][]{Elvis_2004, Risaliti_2005, Puccetti_2007, Turner_2008, Bianchi_2009, Risaliti_2011, Markowitz_2014, Turner_2018}, while several full eclipse events with apparent $ingress$/$egress$ periods have also been captured \citep[e.g.,][]{Lamer_2003, Maiolino_2010, Rivers_2011, Sanfrutos_2013}.      

\par Recently, \citet{Gallo_2021} (hereafter \citetalias{Gallo_2021}) reported a highly distinct occultation event in NGC 6814 captured in 2016 with a high-quality \xmm observation. The $ingress$, eclipse and $egress$ periods are evident in both the count rate (CR) light curves and hardness ratio (HR) curves, while a transiting partial-covering absorber is also revealed with time-resolved spectral fitting analyses. \citetalias{Gallo_2021} interpreted this event as a single Compton-thin cloud eclipsing the central engine; under certain assumptions, properties of the eclipsing cloud (e.g. location, size, density, velocity), along with the size of the X-ray corona, have been derived. Specifically, the eclipsing cloud is derived to be located at a radius of $4.3 \times 10^{10}\,km$ or 2700 $r_{\rm g}$, with diameter $D_{\rm cloud} = 1.3 \times 10^{8}\,km$ or 8.2 $r_{\rm g}$, density $n_{\rm H} = 8.6 \times 10^{9}\,cm^{-3}$, and tangential velocity $v = 10^4\,km\,s^{-1}$, while the diameter of the corona $D_{\rm corona} = 4.2 \times 10^{8}\,km$ or 26 $r_{\rm g}$. 

\par In this work we show this eclipse interpretation could be significantly improved with the aid of the hardness ratio (HR) -- count rate (CR) plot. The HR is commonly defined as (H-S)/(H+S) where H and S are count rates in hard and soft X-ray bands respectively, or equivalently as band ratio S/H (or H/S). The track of an individual source in the HR -- CR plot is determined by two factors, one is the intrinsic spectral variation, and the other is the variation of the absorption. In case of no variable absorption, a ``softer-when-brighter" pattern, i.e., the X-ray spectrum gets softer at higher X-ray fluxes, has been widely seen in AGNs with moderate to high accretion rates, likely driven by the dynamical evolution of the corona \citep[e.g.][]{Wu_2020, Kang_2021, Kang_2022}. The ``softer-when-brighter" pattern often follows a single smooth track for the same source in the HR -- CR plot \citep[e.g.][]{Markowitz_2004,Sobolewska_2009,Soldi_2014,Connolly_2016, Lobban_2020, Kang_2021}, though with subtle or rare deviations \citep{Sarma2015, Wu_2020}, thus searching for irregular HR -- CR diagrams could be a viable approach to find varying absorption or occultation events \citep[e.g.][]{Turner_2018, Cox_2023}, with no need for complicated spectral modelling. In this work we employ the technique reversely to examine the spectral fitting results: if the transient absorption is properly modeled and corrected, the obtained intrinsic HR -- CR diagram should look normal. 
We finally establish a harmonious scenario of the whole occultation event, after comprehensively taking into account the eclipsing absorber and the warm absorber(s), the light curves, the HR -- CR plot, the time-resolved spectra, and two additional \xmm exposures obtained before (2009) and after (2021) the eclipsed exposure.

\section{Data and Reduction} \label{sec:data}

\begin{table}
	\centering
	\caption{\xmm observation logs. }\label{tab:log}
	\renewcommand\tabcolsep{3.5pt}
\begin{tabular}{ccccc} 
\hline
ID & Obs. time & PN Exp (ks) &  PN mode & Filter\\
\hline
0550451801 & 2009-04-22 & 30 & full frame & medium \\
0764620101 & 2016-04-08 & 128 & large window & medium \\
0885090101 & 2021-10-01 & 122 & large window & medium \\
\hline
\end{tabular}
\end{table}

\par To date, three effective \xmm observations of NGC 6814 have been obtained (see Table \ref{tab:log}). The occultation event occurred during the year 2016 exposure, and we refer the other two observations (2009 and 2021) as $former$ and $latter$ in this work. Following \citetalias{Gallo_2021}, we focus on the EPIC-pn data \citep{Struder_2001_PN}, but also include the RGS data \citep{Herder_2001_RGS} to help constrain the ionized absorber(s). 

\par Raw data are reduced using the latest \xmm Science Analysis System (SAS, version 20.0.0) and the Current Calibration Files (CCF). We filter out the intervals suffering from background flares. For EPIC-pn data, the source light curves and spectra are extracted within a circular region with a radius of 60\arcsec , while background from nearby source-free regions. The pile-up effect is found to be negligible with the SAS task \textit{epatplot}. Light curves, with a time bin of 1000 s, are further reprocessed with the task \textit{epiclccorr}, applying background subtraction as well as both absolute and relative corrections, ensuring the light curves of different observations could be directly compared. The EPIC-pn spectra are rebinned to achieve a minimum of 50 counts bin$^{-1}$ using the task \textit{specgroup}. The RGS spectra are extracted using the task \textit{rgsproc}, with a source extraction region including 95\% of point-source events ($xpsfincl=95$). We adopt the first-order spectra and combine those of RGS1 and RGS2 using the task \textit{rgscombine}. 

\par
\begin{figure}
\centering
\subfloat{\includegraphics[width=0.5\textwidth]{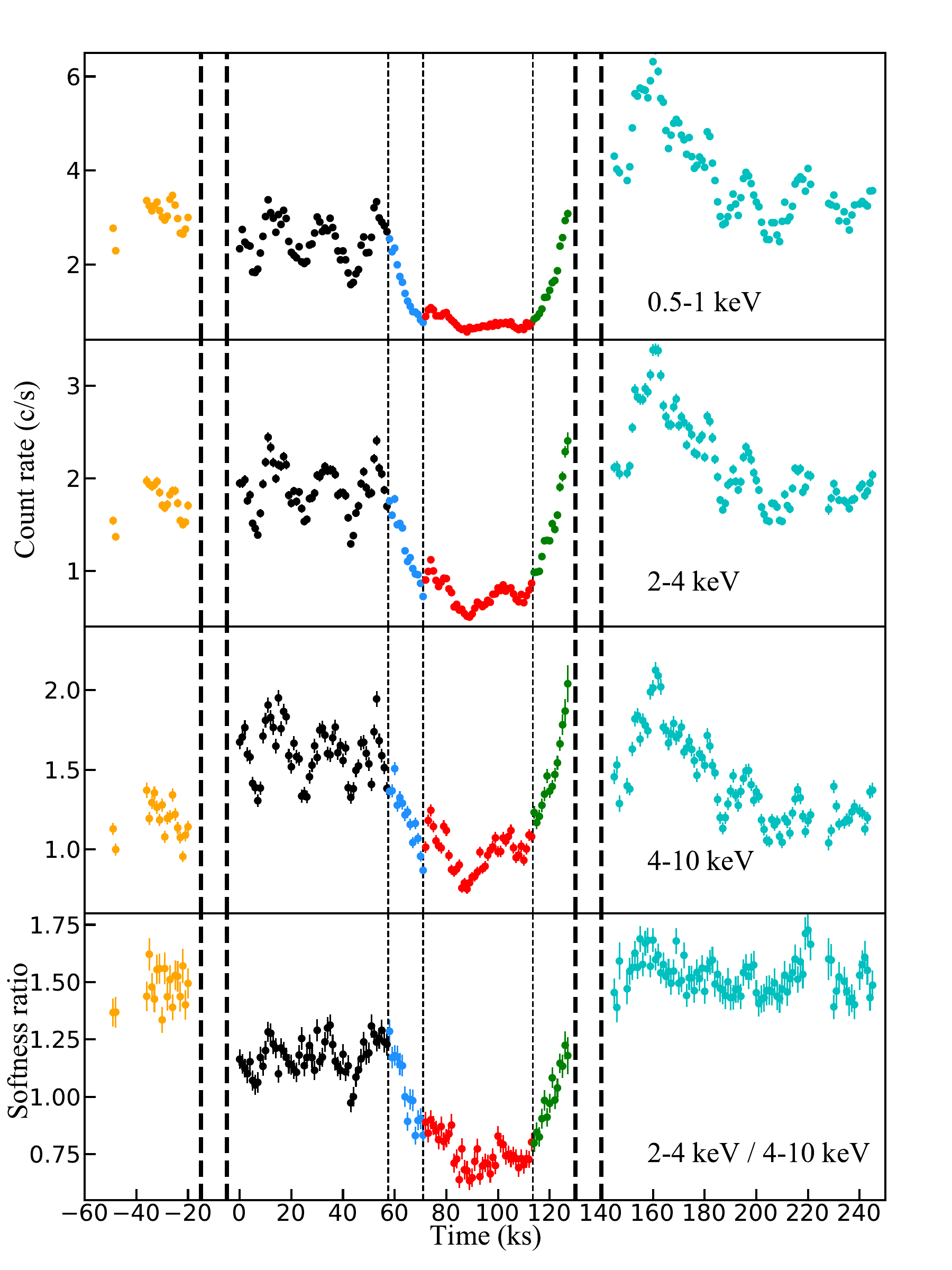}}
\caption{\label{fig:LC} From top to bottom: the 0.5--1 keV, 2--4 keV, 4--10 keV CR curves of EPIC-pn, and the hardness ratio (2--4 keV / 4--10 keV) curve. The $former$, $high$, $ingress$, $low$, $egress$ and $latter$ periods are plotted in orange, black, blue, red, green and cyan respectively (such color configuration is adopted throughout the paper). The bold vertical dashed lines mark the positions where the $x$-axis is discontinuous because of long gaps between three XMM observations.}
\end{figure}

 For the eclipsed exposure obtained in 2016, we also extract the time-resolved spectra\footnote{Note the hidden biases of time-resolved X-ray spectroscopy introduced by \cite{Kang2023} (i.e., splitting light curves horizontally into high/low states) are not applicable here.} by dividing the exposure into four intervals ($high$, $ingress$, $low$, $egress$) following \citetalias{Gallo_2021}. Specifically, the time intervals in kiloseconds are (0, 58), (58, 71), (71, 114) and (114, 125), for the $high$, $ingress$, $low$ and $egress$ periods, respectively (see Fig. \ref{fig:LC}). Time-resolved RGS spectra are also extracted, but only for the $high$ and $low$ periods, as the spectral S/N are much lower for the much shorter $ingress$ and $egress$ periods.

\section{Spectral fitting with the aid of the HR -- CR plot}

\subsection{Warm absorbers revealed in the RGS spectra}

\begin{figure*}
\centering
\subfloat{\includegraphics[width=0.25\textwidth]{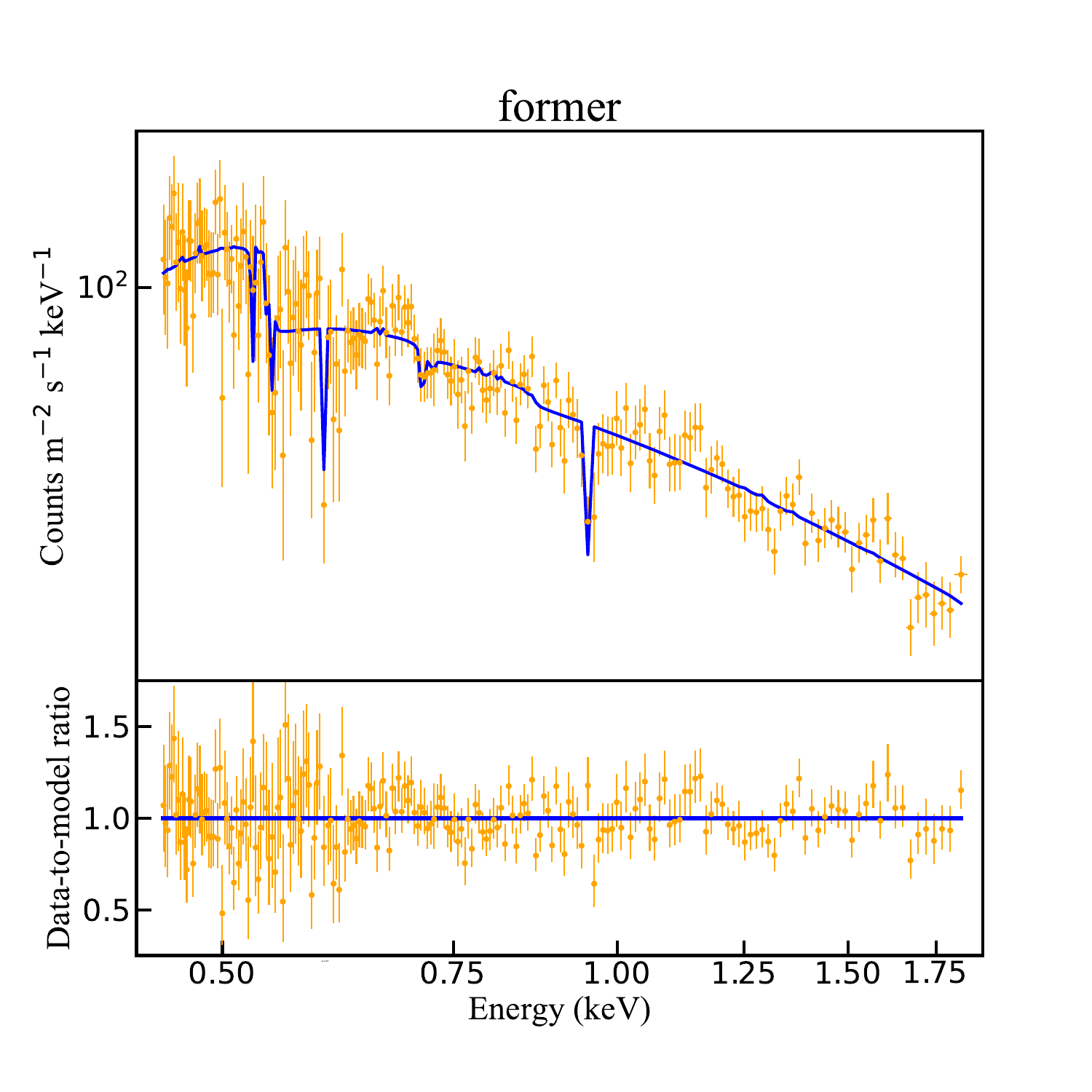}}
\subfloat{\includegraphics[width=0.25\textwidth]{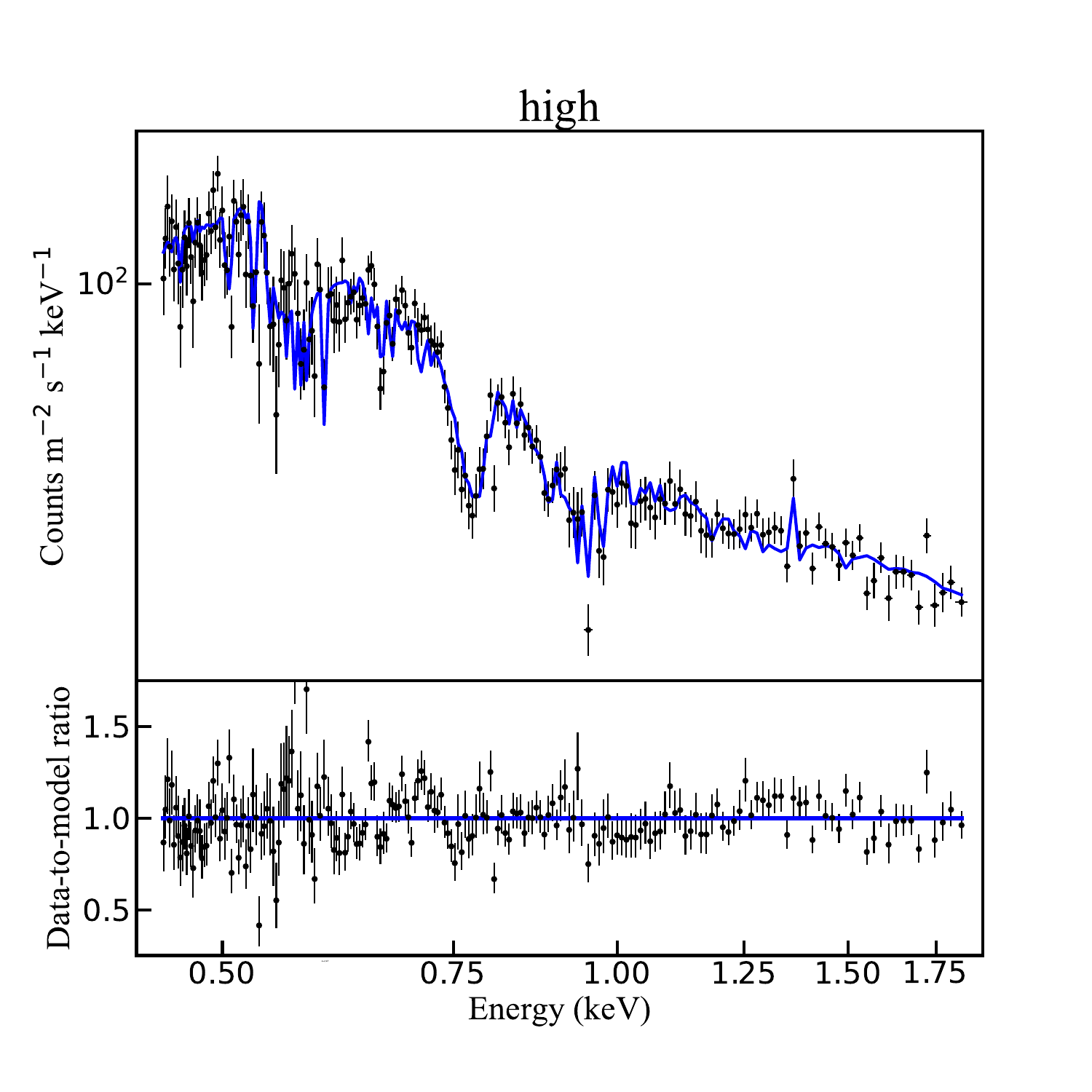}}
\subfloat{\includegraphics[width=0.25\textwidth]{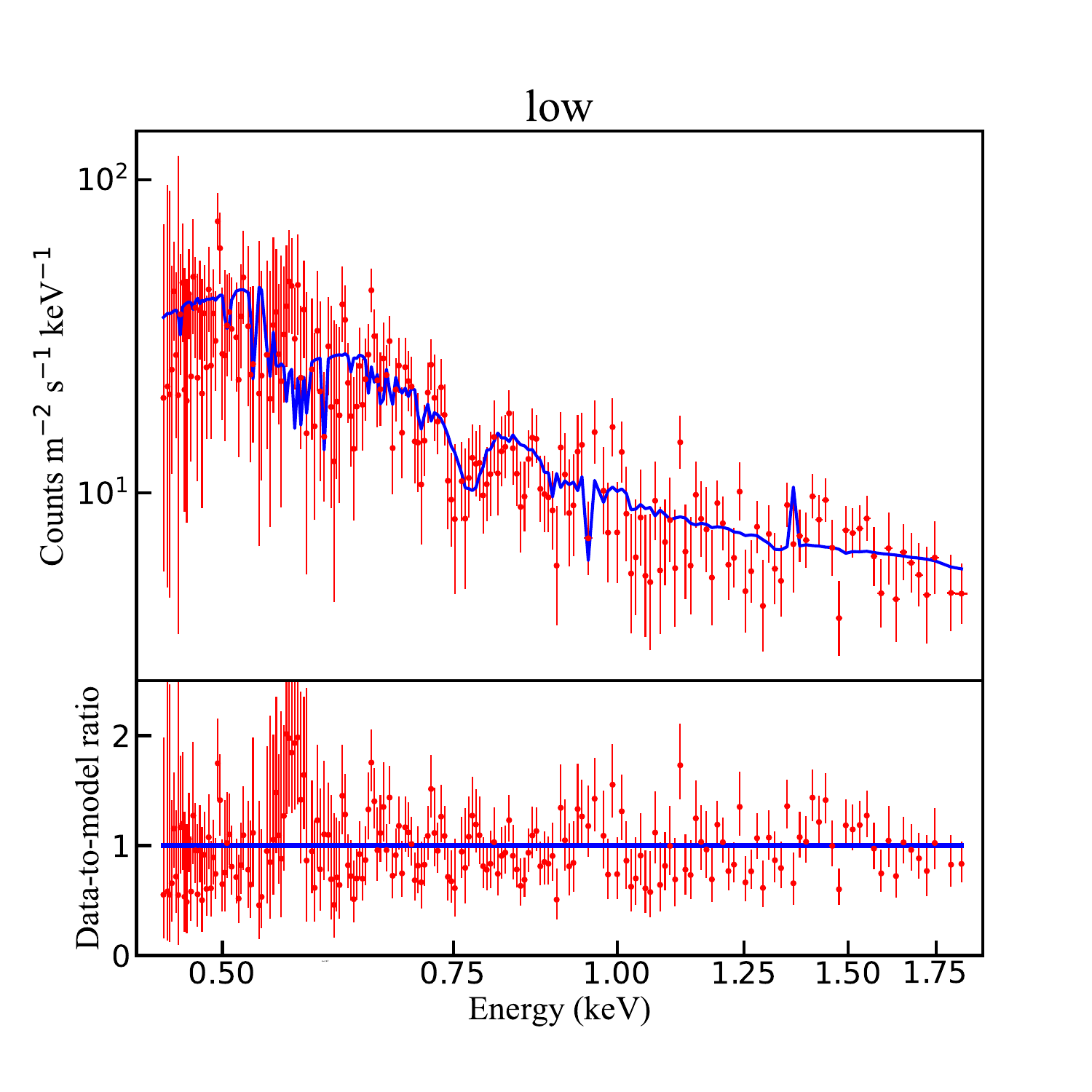}}
\subfloat{\includegraphics[width=0.25\textwidth]{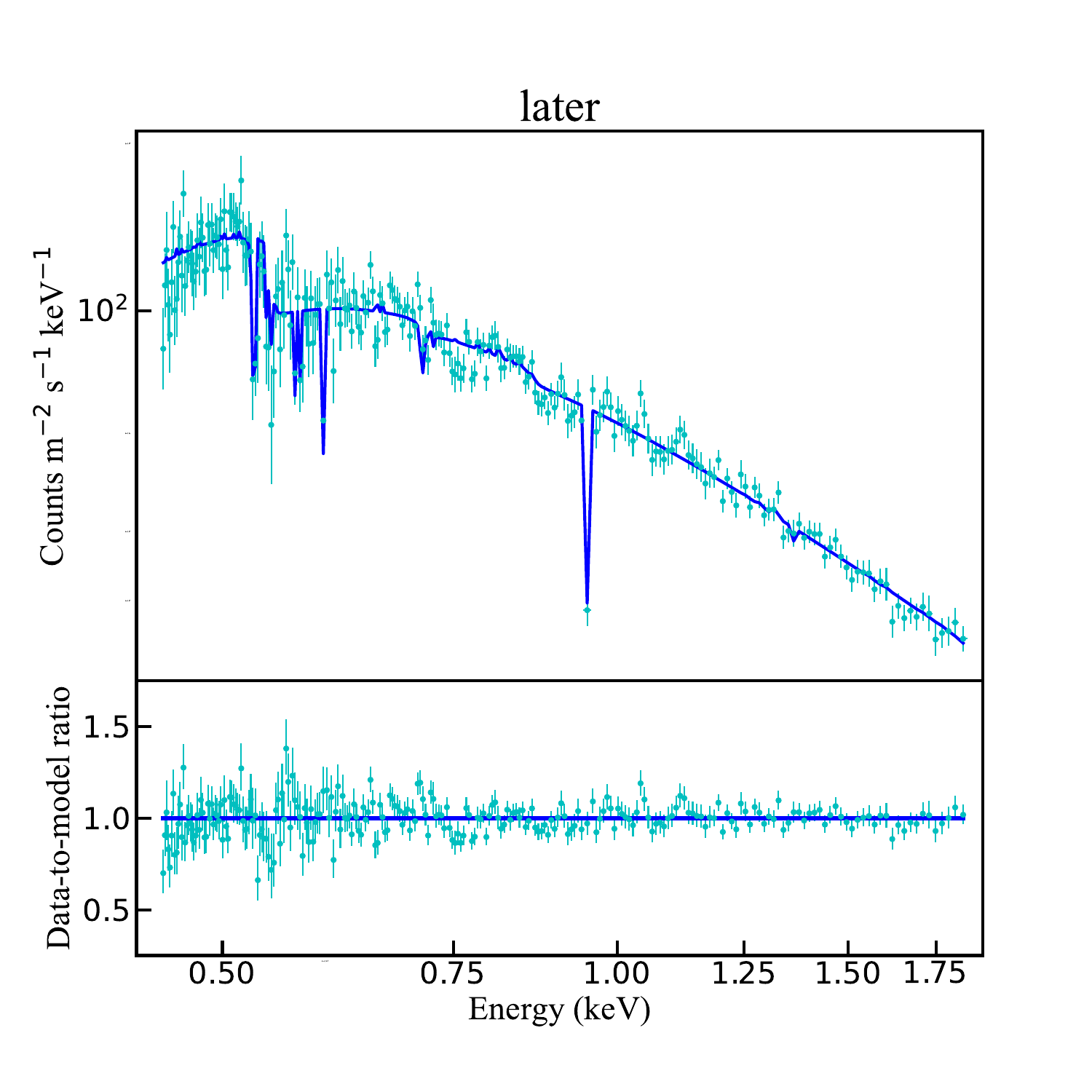}}
\caption{\label{fig:RGS} The RGS spectra and the best-fit models (blue curves), along with the data-to-model ratios, of the $former$, $high$, $low$ and $latter$ exposures. Spectra are rebinned for better illustration.}
\end{figure*}

\begin{table*}

	\centering
	\caption{The spectral fitting results of the RGS spectra$\,^{\rm a}$. }\label{tab:RGS}
	\renewcommand\tabcolsep{10.0pt}
\begin{tabular}{cccccccc} 
\hline
Parameter & $Former$ & $Later$ & $High$ & $Low$  &  $High$ & $Low$  \\
\hline
$\Gamma$ & $1.7^{+0.1}_{-0.1}$  &  $1.9^{+0.1}_{-0.1}$  & $1.3^{+0.2}_{-0.2}$ & $1.3^{+0.5}_{-0.6}$ & $1.3^{+0.2}_{-0.3}$ & $1.0^{+0.4}_{-0.5}$ \\ 
$kT_{\rm bb}$/keV &  $0.09^{+0.01}_{-0.01}$ & $0.10^{+0.01}_{-0.01}$ & $0.12^{+0.01}_{-0.01}$ &
 $0.12^{+0.01}_{-0.01}$ & $0.12^{+0.01}_{-0.01}$ & $0.11^{+0.01}_{-0.01}$ \\
\hline
& & & \multicolumn{2}{c}{$xabs1$} & \multicolumn{2}{c}{$xabs1$} \\
\hline
$N_{\rm H}/\rm 10^{22}\,cm^{-2}$ & - & - & \multicolumn{2}{c}{$0.37^{+0.04}_{-0.04}$} & $0.38^{+0.04}_{-0.04}$ &  $0.28^{+0.09}_{-0.09}$ \\
log $\xi/\rm erg\,cm\,s^{-1}$ & - & - & \multicolumn{2}{c}{$1.03^{+0.05}_{-0.05}$} & $1.05^{+0.05}_{-0.05}$ &   $0.96^{+0.16}_{-0.15}$ \\
$v_{\rm outflow}/ km\, s^{-1}$ & - & - & \multicolumn{2}{c}{ $-5505^{+1246}_{-159}$ } & \multicolumn{2}{c}{ $-4491^{+156}_{-324}$} \\
\hline
& & & \multicolumn{2}{c}{$xabs2$} & \multicolumn{2}{c}{$xabs2$} \\
\hline
$N_{\rm H}/\rm 10^{22}\,cm^{-2}$ & - & - & \multicolumn{2}{c}{$1.48^{+0.65}_{-0.45}$} & $1.68^{+0.75}_{-0.48}$  &  $< 1.80$\\
log $\xi/\rm erg\,cm\,s^{-1}$ & - & - & \multicolumn{2}{c}{$2.78^{+0.09}_{-0.10}$} & $2.74^{+0.10}_{-0.10}$ & $< 5\,^{\rm b}$\\
$v_{\rm outflow}/ km\, s^{-1}$ & - & - & \multicolumn{2}{c}{  $-5419^{+192}_{-267}$  } & \multicolumn{2}{c}{ $-5644^{+137}_{-159}$ } \\
\hline
C/C$_{\rm exp}\,^{\rm c}$ & 2231/2128 & 2247/2056 & \multicolumn{2}{c}{5178/4335} & \multicolumn{2}{c}{5157/4335}\\
\hline
\end{tabular}
\begin{tablenotes}
\item {a: For the $high$ and $low$ states, we show the fitting results based on two different settings of parameter coupling. The intermediate column shows the results where the parameters of the warm absorbers ($N_{\rm H}$, log $\xi$, $v_{\rm outflow}$) are linked between the $high$ and $low$ states, while the right column shows the case where $N_{\rm H}$, log $\xi$ are free to vary between two states.}
\item {b: Can not be constrained within the hard limit of the model. } 
\item {c: The expected C is calculated following \citet{Kaastra_2017}. } 
\end{tablenotes}
\end{table*}

\par We first note that in this work we fit RGS and EPIC-pn spectra separately. Performing joint-fitting of them is infeasible, because the RGS spectrum contains much fewer counts than the EPIC-pn spectrum, making the RGS spectrum statistically insignificant during joint-fitting. Moreover, the covering fraction of the eclipsing absorber is rapidly varying during the observation, which however could barely be constrained with the limited band coverage of the RGS spectra.

\par In this work the spectral analyses are conducted as described below. Firstly we fit the RGS spectra with SPEX \citep{Kaastra_1996, Kaastra_2020} to detect whether warm absorbers exist. If existing, the corresponding warm absorbers will be included when fitting the EPIC-pn spectra with XSPEC \citep{Arnaud_1996}. During fitting, the column density and ionization parameter are free to vary, while the outflow velocity is frozen at the best-fit result of the RGS spectra. Then the best-fit absorber(s) from the EPIC-pn spectra (including a partially covering absorber in the eclipsing observation) will be fed back to fit the RGS spectra, and we will confirm that the fitting of the RGS spectra does not seriously deteriorate. In summary, the RGS spectra assist by determining the warm absorbers and double-checking the best-fit results of EPIC-pn spectra. 

\par The fitting of RGS spectra is carried out in 0.45 -- 1.85 keV with SPEX, using C-statistics \citep{Cash_1979, Kaastra_2017} and abundances given by \citet{ANDERS_1989}. We show the spectra of the four intervals ($former$, $high$, $low$ and $latter$) in Fig. \ref{fig:RGS}. One could tell at a glance that the spectra of the $former$ and $latter$ exposures are rather plain (after ignoring the features of the RGS response), while the spectra of the eclipsed observation ($high$ and $low$) show a prominent and wide absorption feature around 0.75 keV \citep[namely the iron unidentified transition array, UTA, ][]{Sako_2001,Behar_2001}. We then adopt a power law plus a black body to fit the intrinsic continuum, and include the Galactic absorption of $N_{\rm H} = \rm 1.5 \times 10^{21}\,cm^{-2}$ \citep{Willingale_2013} using the model \textit{hot} \citep{dePlaa_2004}. The spectra of the $former$ and $latter$ observations are well fitted with this simple model, and additional warm absorber(s) are statistically not required (see \S \ref{sec:4.1.2} for further discussion on these warm absorbers). Meanwhile, the joint-fitting of the $high$ and $low$ spectra with the simple continuum model yields large residuals, likely caused by the warm absorber(s) reported in \citetalias{Gallo_2021}. Two $xabs$ components \citep{Steenbrugge2003} are thus included to model the warm absorbers in the spectra, which significantly improve the spectral fitting ($\Delta$C = -250 for the first component, and $\Delta$C = -60 for the second one, when performing the joint-fitting for the two states). Moreover, allowing the $N_{\rm H}$ and log $\xi$ free to vary between $high$ and $low$ spectra slightly improves the fitting ($\Delta$C = -21), showing the warm absorbers are weaker in the $low$ period (as shown below, the variation of the warm absorbers is also statistically required while fitting the EPIC-pn spectra). The fitting results to RGS spectra are shown in Table \ref{tab:RGS}, with errors and upper/lower limits derived following $\Delta \chi^2=2.71$ criteria, corresponding to the 90 per cent confidence level for one interesting parameter. Note the two absorbers have quite different $N_{\rm H}$ and log $\xi$, but a similar outflowing velocity $v_{\rm outflow} \sim 5500 \, km \, s^{-1}$. Note in literature the phrase ``warm absorber" generally refers to slowly outflowing ionized absorbers with typical $v < 1000 \, km \, s^{-1}$ \citep[e.g., ][]{Laha_2014,Gallo_2023}, located far from the black hole. In this work, following \citetalias{Gallo_2021}, we refer the two ionized absorbers identified in the RGS spectrum as warm absorbers, regardless of their large $v_{\rm outflow}$, but see the end of \S\ref{sec:4.1.2} for further discussion.  

\par In conclusion, we find remarkable ionized absorption features during the eclipsed observation, which are however absent in the $former$ and $latter$ observations. From this perspective, the $high$ period, before the distinct drop of the flux, is actually already eclipsed compared with the other two \xmm exposures, at least by some warm absorbers (see Fig. \ref{fig:RGS} for the prominent difference between the $high$ period and other two exposures). This is further supported by the HR curves in Fig. \ref{fig:LC}, where even the $high$ interval (black points) has significantly softer spectrum than the $former$ and $latter$ one (orange and cyan points). Meanwhile, spectral fitting suggests weaker warm absorption in the $low$ period compared with the $high$ one, though only at a moderate confidence due to the poor quality of the $low$ spectrum. 

\subsection{Testifying the eclipse model of \citetalias{Gallo_2021} with the HR -- CR diagram}

\begin{figure*}
\centering
\subfloat{\includegraphics[width=0.33\textwidth]{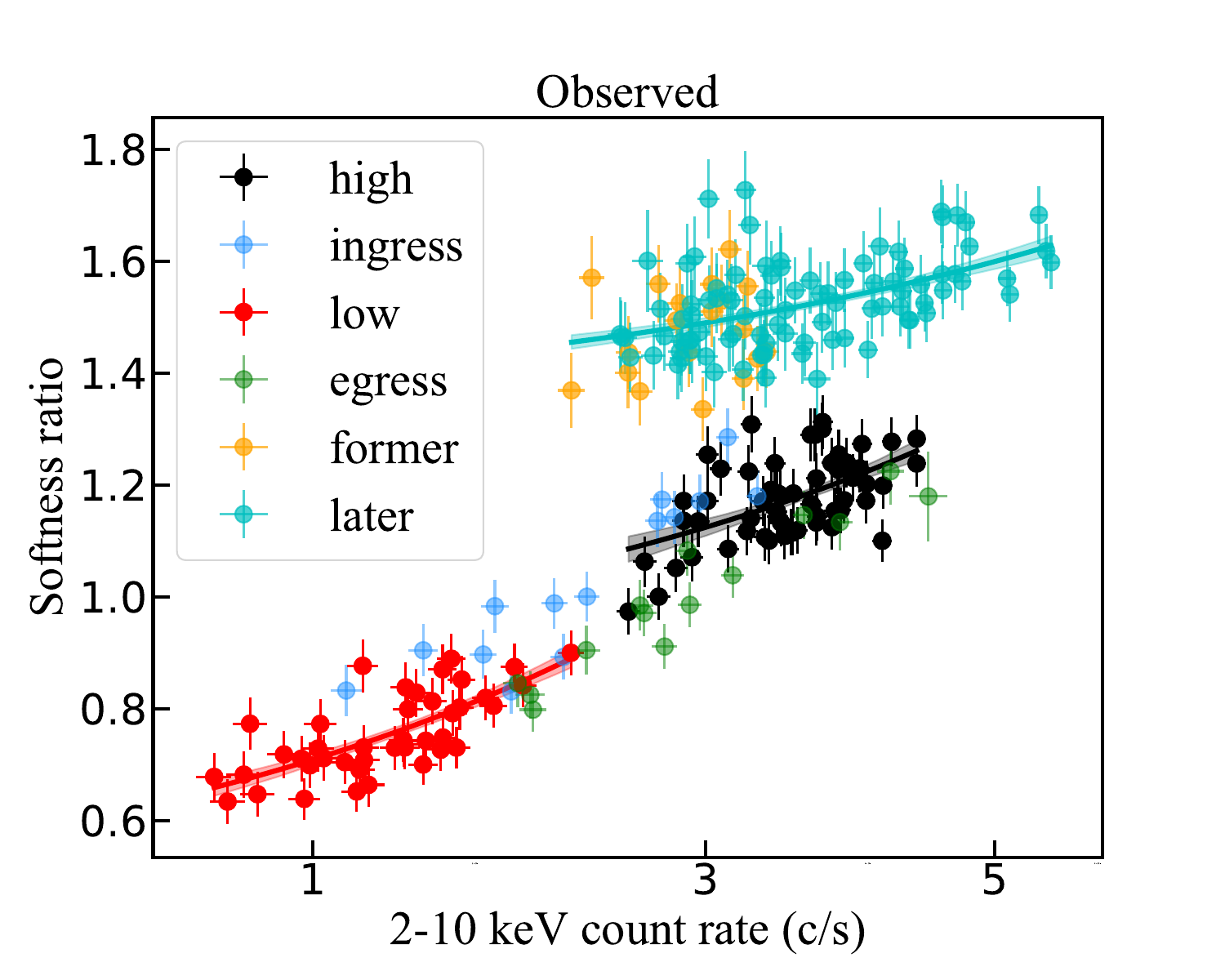}}
\subfloat{\includegraphics[width=0.33\textwidth]{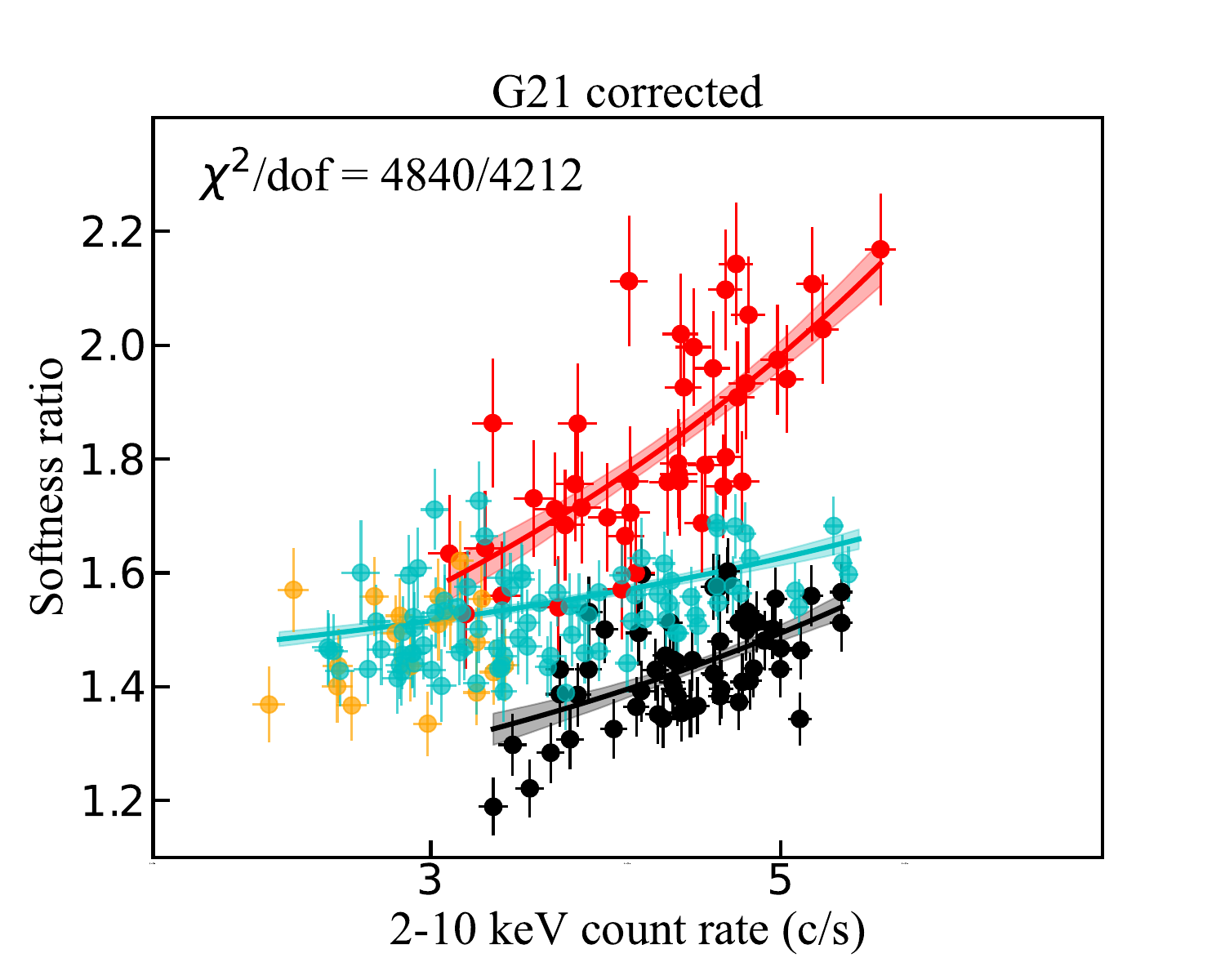}}
\subfloat{\includegraphics[width=0.33\textwidth]{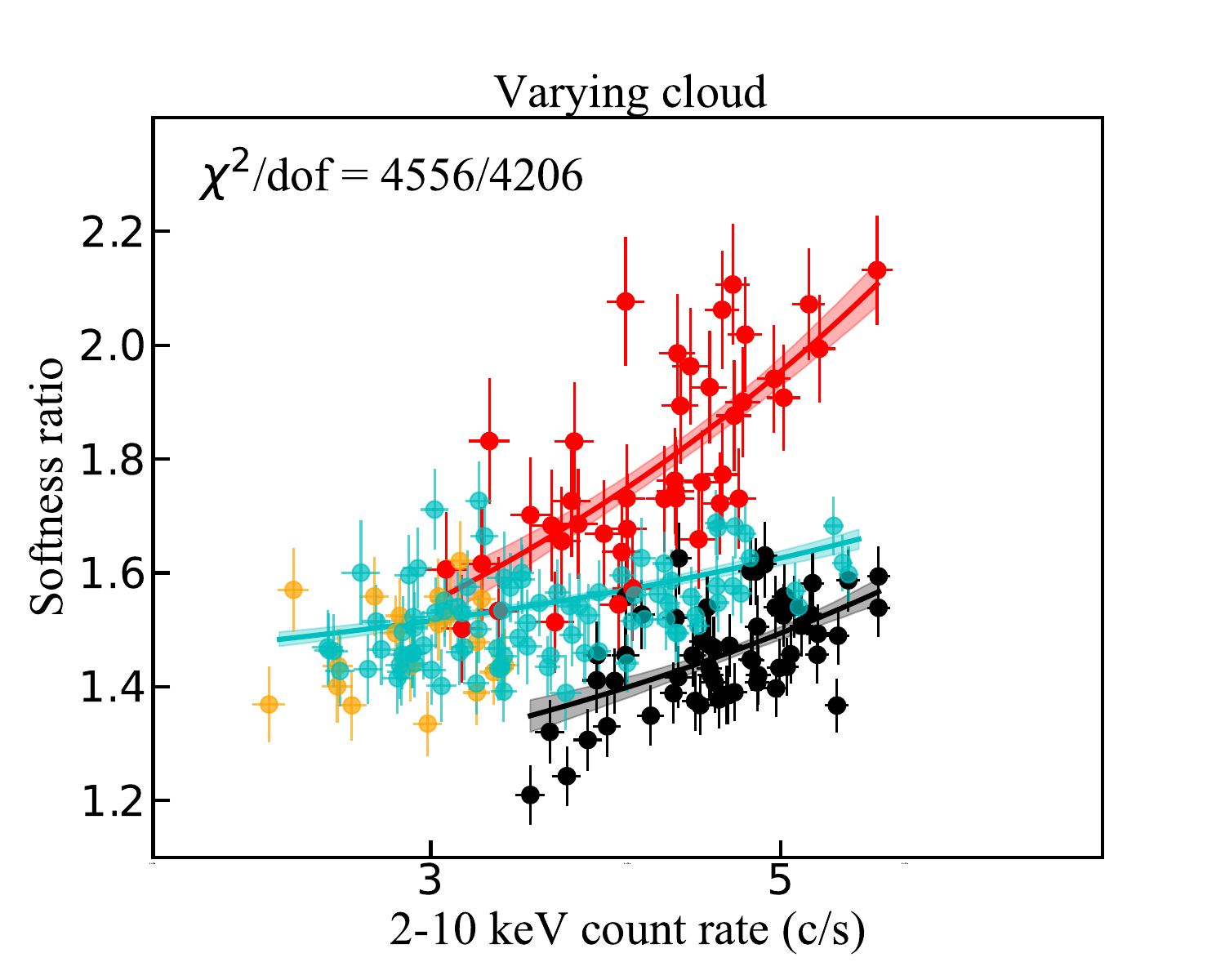}}\\
\subfloat{\includegraphics[width=0.33\textwidth]{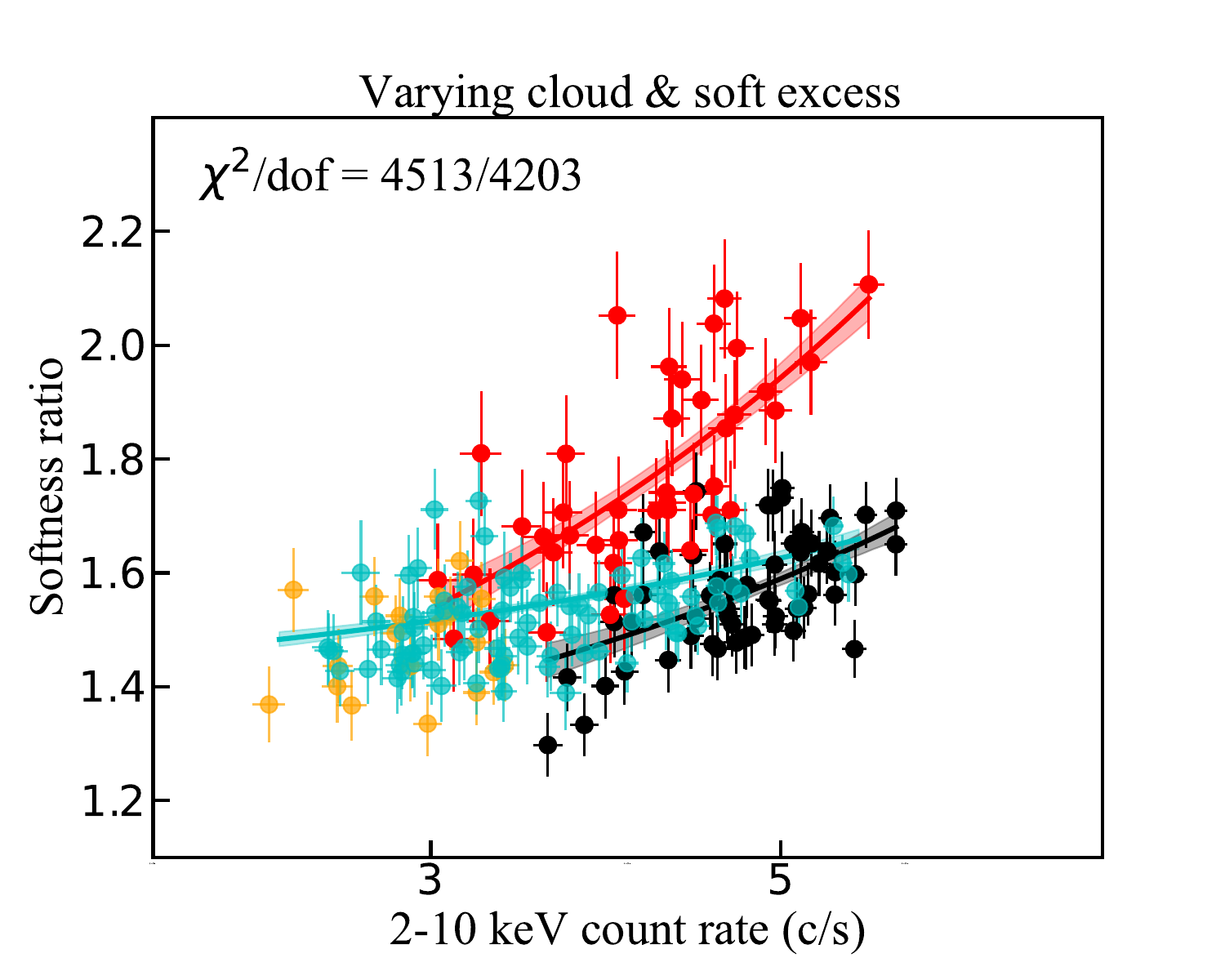}}
\subfloat{\includegraphics[width=0.33\textwidth]{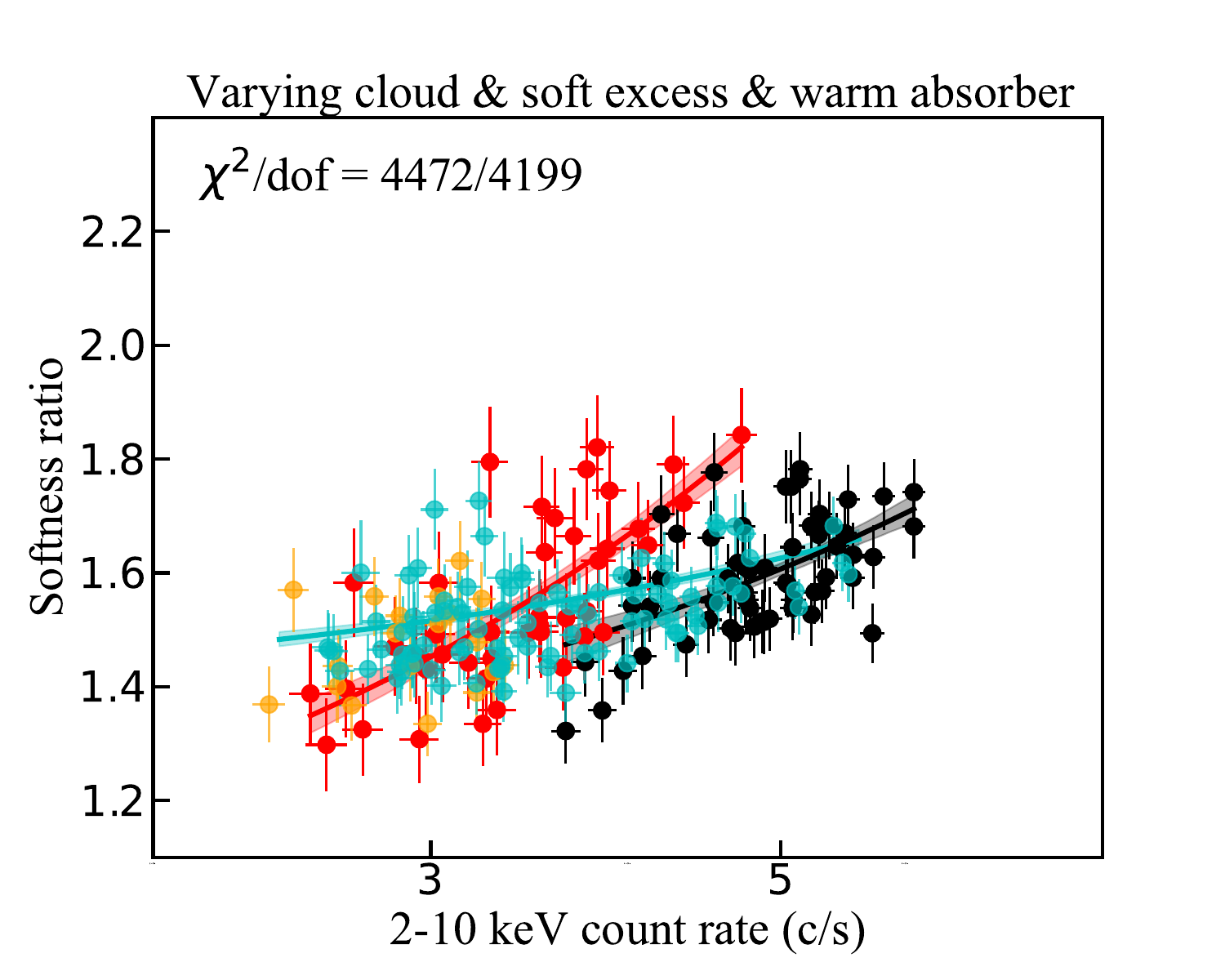}}
\subfloat{\includegraphics[width=0.33\textwidth]{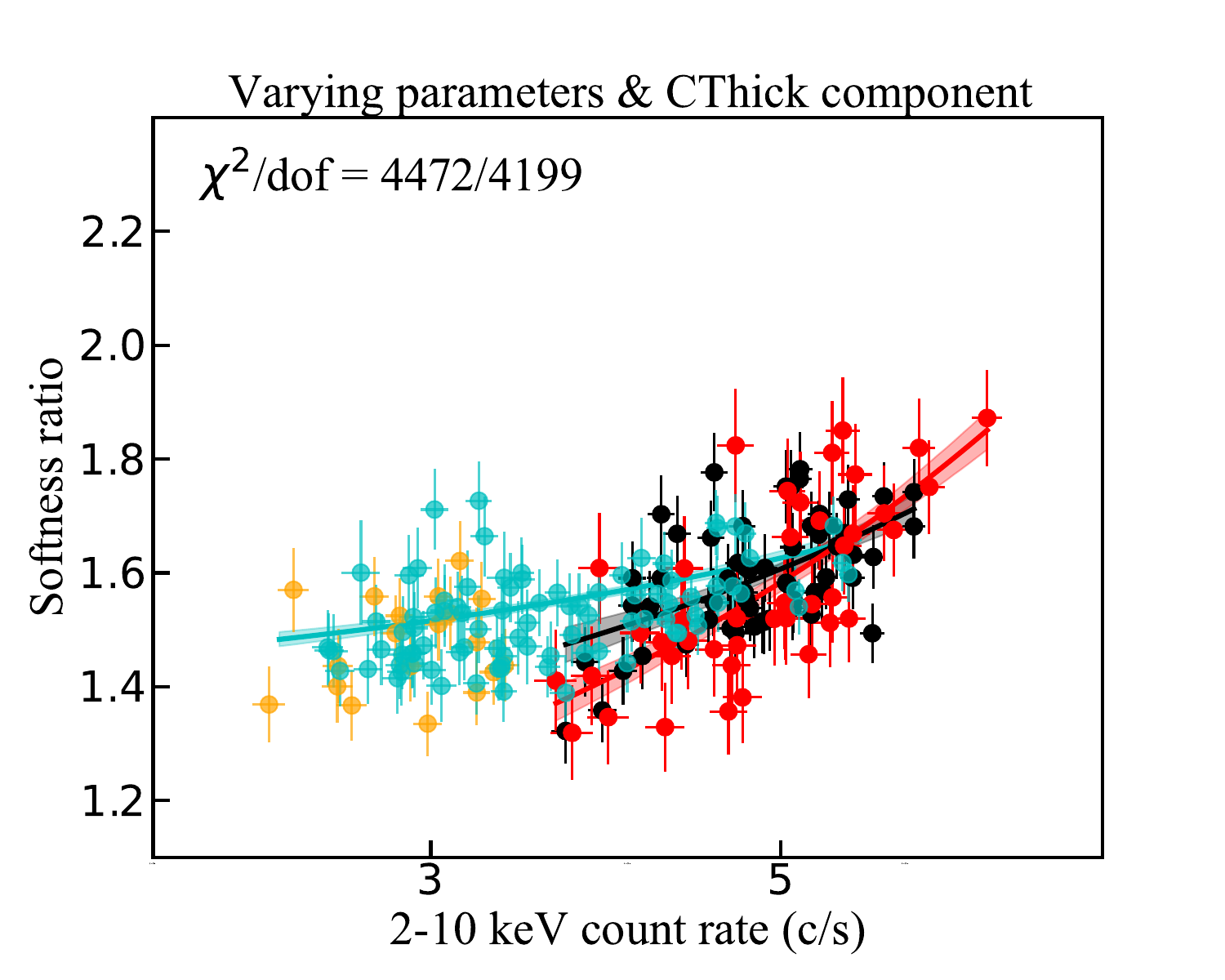}}
\caption{\label{fig:SR} The top left panel shows the observed hardness ratio (2--4 keV/4--10 keV) versus the 2--10 keV count rate. The other five panels show the HR -- CR diagrams after the absorption being corrected, based on a series of progressively improved models (see the text for the detailed correction approach and models). Note except for the upper left panel, all other panels have identical coordinate ranges. Three lines in each panel plot the best-fit linear regression (with 1$\sigma$ confidence range) between HR and CR for the $former$+$latter$ exposures (cyan), the $high$ (black) and $low$ period (red), respectively, to highlight their tracks in the HR -- CR diagram. 
}
\end{figure*}  

\par In Fig. \ref{fig:SR} we plot the hardness ratio (2--4 keV/4--10 keV) versus the 2--10 keV count rate using the 1000s bin light curves presented in Fig. \ref{fig:LC} (top left panel), where the bands are selected to avoid the complexity of the soft excess. 
A clear ``softer-when-brighter" trend is seen both in the $latter$ exposure, and the 2016 exposure when the eclipse event was captured. While the $former$ exposure appears following the same track with the $latter$ one in the HR -- CR plot, however, the ``softer-when-brighter" trend revealed by the 2016 exposure is considerably steeper and along an offset track.

\par We then fit the EPIC-pn spectra in the 0.5--10 keV band using XSPEC and $\chi^2$ statistics, adopting the spectral model of \citetalias{Gallo_2021} which
consists of invariable absorbers (warm absorbers and Galactic absorption), an eclipsing absorber, a primary power law continuum, a soft excess, and a Fe K$\alpha$ line.
We simplify the model of \citetalias{Gallo_2021} by replacing the two $nthComp$ with $bbody$ and $powerlaw$. These models fit the continuum as good as the Comptonization model in the concerned band and provide more intuitive parameters that can be directly compared with the fitting results of the RGS spectra. In addition, instead of using the XSPEC table given by \citet{Parker_2019}, we create our own $xabs$ table, ensuring that the abundance and fixed parameters are the same, so that a direct comparison between the results of SPEX and XSPEC is feasible. Specifically, $N_{\rm H}$ and log $\xi$ are variable parameters, $v_{\rm outflow}$ is fixed the best-fit value in Table \ref{tab:RGS}, and other parameters not shown in Table \ref{tab:RGS} (e.g. broadening velocity and elemental abundances) are fixed at the default values of $xabs$.

\par Intrinsic absorption however is statistically not required while fitting the EPIC-pn spectra of the $former$ and $latter$ observations\footnote{The $latter$ observation has a coordinated NuSTAR exposure, in which we find no significant absorption either.}, thus $TBabs \times  (bbody + powerlaw + zgauss)$ is adopted (see Table \ref{tab:PNFL} for the fitting results). In other words, the observed spectra and HR -- CR diagram of these two observations are substantially intrinsic, suitable as a reference for the 2016 exposure.
For the eclipsed exposure, the adopted model is $TBabs \times (xabs \times xabs \times zxipcf \times (bbody + powerlaw) + zgauss)$, where $TBabs$ accounts for the Galactic absorption, $zxipcf$ for the eclipsing absorber, $zgauss$ for the Fe K$\alpha$ line. The spectra of the four periods are jointly fitted with the model parameters linked as \citetalias{Gallo_2021}. Our independent fitting with the \citetalias{Gallo_2021} model does yield results (see Table \ref{tab:results}) consistent with \citetalias{Gallo_2021} and successfully reveal a scenario of occultation.

\begin{table}
 \footnotesize
	\centering
	\caption{The spectral fitting results of the EPIC-pn spectra of the $former$ and $latter$ observation. }\label{tab:PNFL}
	\renewcommand\tabcolsep{1.0pt}
\begin{tabular}{cccc} 
\hline
Component & Parameter & $Former$ &  $Later$ \\
\hline
bbody & kT/keV & $0.087^{+0.007}_{-0.009}$ & $0.087^{+0.002}_{-0.002}$ \\
  & log flux/$\rm erg\,cm^{-2}\,s^{-1}$ & $-12.016^{+0.054}_{-0.051}$ & $-11.841^{+0.015}_{-0.016}$ \\
\hline
powerlaw & $\Gamma$ & $1.731^{+0.019}_{-0.012}$ & $1.743^{+0.004}_{-0.004}$ \\
  & log flux/$\rm erg\,cm^{-2}\,s^{-1}$ & $-10.450^{+0.004}_{-0.003}$ & $-10.343^{+0.001}_{-0.001}$ \\
\hline
zgauss & E/keV & $6.45^{+0.08}_{-0.08}$ & $6.44^{+0.02}_{-0.02}$ \\
  & $\sigma$/keV & $0.27^{+0.15}_{-0.19}$ & $0.12^{+0.03}_{-0.02}$ \\
  & $\rm Norm/10^{-5}\,ph.\,cm^{2}\,s^{-1}$ & $4.8^{+1.6}_{-1.9}$ & $3.7^{+0.4}_{-0.4}$ \\
\hline
  & $\chi^2$/dof & 1279/1264 & 1852/1640 \\
\hline
\end{tabular}
\begin{tablenotes}
\item{*}{All the fluxes reported in this paper are the unabsorbed fluxes calculated in 0.5--10 keV band using model $cflux$, unless otherwise noted.}
\end{tablenotes}
\end{table}

\par In the upper middle panel of Fig. \ref{fig:SR}, we present the absorption-corrected HR -- CR plot. For the $former$ and $latter$ observations only the Galactic absorption is corrected as we find no significant absorption. While correcting the absorption slightly reduces the difference between the 2016 exposure and the other two, however, new and prominent deviation between the $high$ and $low$ periods emerges (see the clear offset between the $high$ and $low$ periods in the HR -- CR plot). Note that this practice applies an average correction of the absorption for each period, thus could be inaccurate for the $ingress$ and $egress$ periods during which the absorption is varying rapidly. Therefore, we omit these two periods from the absorption corrected panels in Fig. \ref{fig:SR}. 

\par It is very puzzling that, while the 2016 exposure exhibits roughly a single track in the HR -- CR plot before absorption correction, correcting the transient absorption yields two rather different tracks during one exposure. If the absorption correction is proper, this would mean the intrinsic spectrum of NGC 6814 significantly steepens by coincidence during the eclipse (see also the variation of the best-fit $\Gamma$ in Table \ref{tab:results}), which is however extremely unlikely. Since the eclipse is highly evident in the light curves and in the spectral fitting results, below we revisit the spectral fitting to seek for possible solution(s) within the framework of occultation.

\subsection{Revising the spectral fitting and the absorption model}

\par We note the \citetalias{Gallo_2021} model is constructed on three assumptions, a homogeneous eclipsing cloud, an invariable soft excess, and invariable warm absorbers. However, the eclipsing cloud can have highly asymmetrical geometry and be inhomogeneous \citep[e.g., ][]{Maiolino_2010}. Besides, as shown in Fig. \ref{fig:LC}, the 0.5--1 keV band (where the soft excess contributes to $\sim$ 30\% of the count rate) shows a similar variation pattern to the other bands, indicating it could be over-simplified to assume a constant soft X-ray excess. Meanwhile, our analysis of the RGS spectra implies varying warm absorbers. Below we lift the three assumptions one by one.  

\begin{table*}
	\centering
	\caption{The spectral fitting results of the EPIC-pn spectra of the 2016 exposure.}\label{tab:results}
\begin{tabular}{cccccccc} 
\hline
Component & Parameter & \multicolumn{5}{c}{Value} \\
& & All & $High$ & $Ingress$ & $Low$ & $Egress$ \\
\hline
\hline
\multicolumn{7}{c}{Parameters linked as \citetalias{Gallo_2021}} \\
\hline
$xabs1$ & $N_{\rm H}/\rm 10^{22}\,cm^{-2}$ & $0.46^{+0.01}_{-0.05}$ & & & &\\
 & log $\xi/\rm erg\,cm\,s^{-1}$ & $1.06^{+0.02}_{-0.04}$ & & & &\\
\hline
$xabs2$ & $N_{\rm H}/\rm 10^{22}\,cm^{-2}$ & $2.96^{+0.23}_{-0.29}$ & & & &\\
 & log $\xi/\rm erg\,cm\,s^{-1}$ & $2.97^{+0.03}_{-0.03}$ & & & &\\
\hline
zxipcf & $N_{\rm H}/\rm 10^{22}\,cm^{-2}$ & $13.1^{+0.3}_{-0.7}$ & & & &\\
 & log $\xi/\rm erg\,cm\,s^{-1}$ & $1.09^{+0.02}_{-0.04}$ & & & &\\
 & $f_{\rm cov}$ & & $0.07^{+0.02}_{-0.06}$ & $0.39^{+0.03}_{-0.02}$ & $0.80^{+0.01}_{-0.01}$ & $0.55^{+0.05}_{-0.03}$ \\
\hline
bbody & kT/keV & $0.112^{+0.003}_{-0.001}$ & & & &\\
 & log flux/$\rm erg\,cm^{-2}\,s^{-1}$ & $-11.19^{+0.04}_{-0.02}$ & & & &\\
\hline
powerlaw & $\Gamma$ & & $1.62^{+0.03}_{-0.03}$ & $1.68^{+0.03}_{-0.03}$ & $1.98^{+0.01}_{-0.02}$ & $1.72^{+0.03}_{-0.01}$ \\
 & log flux/$\rm erg\,cm^{-2}\,s^{-1}$ & & $-10.3^{+0.01}_{-0.01}$ & $-10.32^{+0.03}_{-0.01}$ & $-10.24^{+0.01}_{-0.02}$ & $-10.22^{+0.11}_{-0.03}$ \\
\hline
zgauss & E/keV & $6.42^{+0.01}_{-0.01}$ & & & &\\
 & $\sigma/\rm eV$ & $0.15^{+0.02}_{-0.02}$ & & & &\\
 & $\rm Norm/10^{-5}\,ph.\,cm^{2}\,s^{-1}$ & & $6.0^{+0.6}_{-0.5}$ & $5.3^{+0.9}_{-0.9}$ & $5.1^{+0.6}_{-0.3}$ & $5.8^{+1.0}_{-1.0}$ \\
\hline
& $\chi^2$/dof & 4841/4212 & & & &\\
\hline
\hline
\multicolumn{7}{c}{Final model in this work} \\
\hline
$xabs1$ & $N_{\rm H}/\rm 10^{22}\,cm^{-2}$ & & $0.33^{+0.03}_{-0.03}$  & - & $0.22^{+0.04}_{-0.04}$ & - \\
 & log $\xi/\rm erg\,cm\,s^{-1}$ & & $1.14^{+0.05}_{-0.05}$ &  - & $0.84^{+0.13}_{-0.13}$ & - \\
\hline
$xabs2$ & $N_{\rm H}/\rm 10^{22}\,cm^{-2}$ & & $2.55^{+0.42}_{-0.39}$  & - & $0.58^{+0.37}_{-0.23}$ & - \\
 & log $\xi/\rm erg\,cm\,s^{-1}$ & & $3.0^{+0.05}_{-0.04}$  & - & $3.0^{+0.10}_{-0.09}$ & - \\
\hline
zxipcf & $N_{\rm H}/\rm 10^{22}\,cm^{-2}$ & & $12.0^{+0.9}_{-0.8}$ & $11.3^{+1.2}_{-2.0}$ & $17.2^{+0.6}_{-1.4}$ & $12.4^{+2.6}_{-3.2}$ \\
 & log $\xi/\rm erg\,cm\,s^{-1}$ & & $1.98^{+0.07}_{-0.06}$ & $1.91^{+0.08}_{-0.19}$ & $1.91^{+0.02}_{-0.31}$ & $1.67^{+0.28}_{-0.23}$ \\
 & $f_{\rm cov}$ & & $0.38^{+0.04}_{-0.04}$ & $0.44^{+0.07}_{-0.08}$ & $0.76^{+0.04}_{-0.02}$ & $0.55^{+0.06}_{-0.08}$ \\
\hline
pcfabs & $N_{\rm H}/\rm 10^{22}\,cm^{-2}$ & 500 & & & & \\
 & $f_{\rm cov}$ & & 0 & 0.15 & 0.30 & 0.15 \\
\hline
bbody & kT/keV & $0.102^{+0.003}_{-0.003}$ & & & &\\
 & log flux/$\rm erg\,cm^{-2}\,s^{-1}$ & & $-11.29^{+0.03}_{-0.03}$ & $-11.32^{+0.04}_{-0.04}$ & $-11.29^{+0.07}_{-0.06}$ & $-11.37^{+0.05}_{-0.06}$ \\
\hline
powerlaw & $\Gamma$ & & $1.76^{+0.03}_{-0.04}$ & $1.71^{+0.07}_{-0.07}$ & $1.73^{+0.11}_{-0.05}$ & $1.69^{+0.08}_{-0.09}$ \\
 & log flux/$\rm erg\,cm^{-2}\,s^{-1}$ & & $-10.22^{+0.02}_{-0.02}$ & $-10.28^{+0.04}_{-0.04}$ & $-10.22^{+0.06}_{-0.02}$ & $-10.22^{+0.04}_{-0.05}$ \\
\hline
zgauss & E/keV & $6.41^{+0.01}_{-0.01}$ & & & &\\
 & $\sigma/\rm eV$ & $0.13^{+0.02}_{-0.02}$ & & & &\\
 & $\rm Norm/10^{-5}\,ph.\,cm^{2}\,s^{-1}$ & & $5.1^{+0.6}_{-0.5}$ & $5.6^{+0.9}_{-0.9}$ & $4.5^{+0.5}_{-0.5}$ & $6.2^{+1.1}_{-1.1}$ \\
\hline
& $\chi^2$/dof & 4472/4199 & & & &\\
\hline
\end{tabular}
\begin{tablenotes}
\item{In our final model, the outflow velocity of the warm absorbers is set at the best-fit value of the RGS spectra. $N_{\rm H}$ and log $\xi$ of the warm absorbers are free to vary in the $low$ period, while linked together among the $high$, $ingress$ and $egress$ periods. }
\end{tablenotes}
\end{table*}

\par We first untie the parameters of the eclipsing cloud (column density $N_{\rm H}$ and ionization parameter log $\xi$), allowing them to vary between periods during the 2016 exposure. We find a larger $N_{\rm H}$ and smaller log $\xi$ in the $low$ period than in the $high$ period. The fitting is significantly improved with $\Delta \chi^2 = -284$ and F-test giving a probability $< 10^{-51}$. However, the discrepancy in the HR -- CR remains (see the top right panel in Fig. \ref{fig:SR}). We further allow the normalization of the soft excess to vary among periods. The soft excess is found to be weaker in the $low$ period, which is also statistically significant, with $\Delta \chi^2 = -43$ and F-test giving a probability $\sim 1.1 \times 10^{-8}$. On this occasion, the $high$ period becomes more consistent (but with slightly steeper slope) with the $former$ and $latter$ observations in the HR -- CR diagram, while the $low$ period still deviates from the others (lower left panel in Fig. \ref{fig:SR}). Finally, we investigate the case of varying warm absorbers. Considering the poor quality of the spectra in the $ingress$ and $egress$ periods, we only untie the warm absorbers ($xabs$ components) of the $low$ period, leaving parameters of the $high$, $ingress$ and $egress$ periods linked during fitting. We obtain consistent results with those of the RGS spectra, that the warm absorbers become weak during the $low$ period, with $\Delta \chi^2 = -41$ and F-test giving a probability $\sim 1.0 \times 10^{-7}$.

\begin{figure}
\centering
\subfloat{\includegraphics[width=0.45\textwidth]{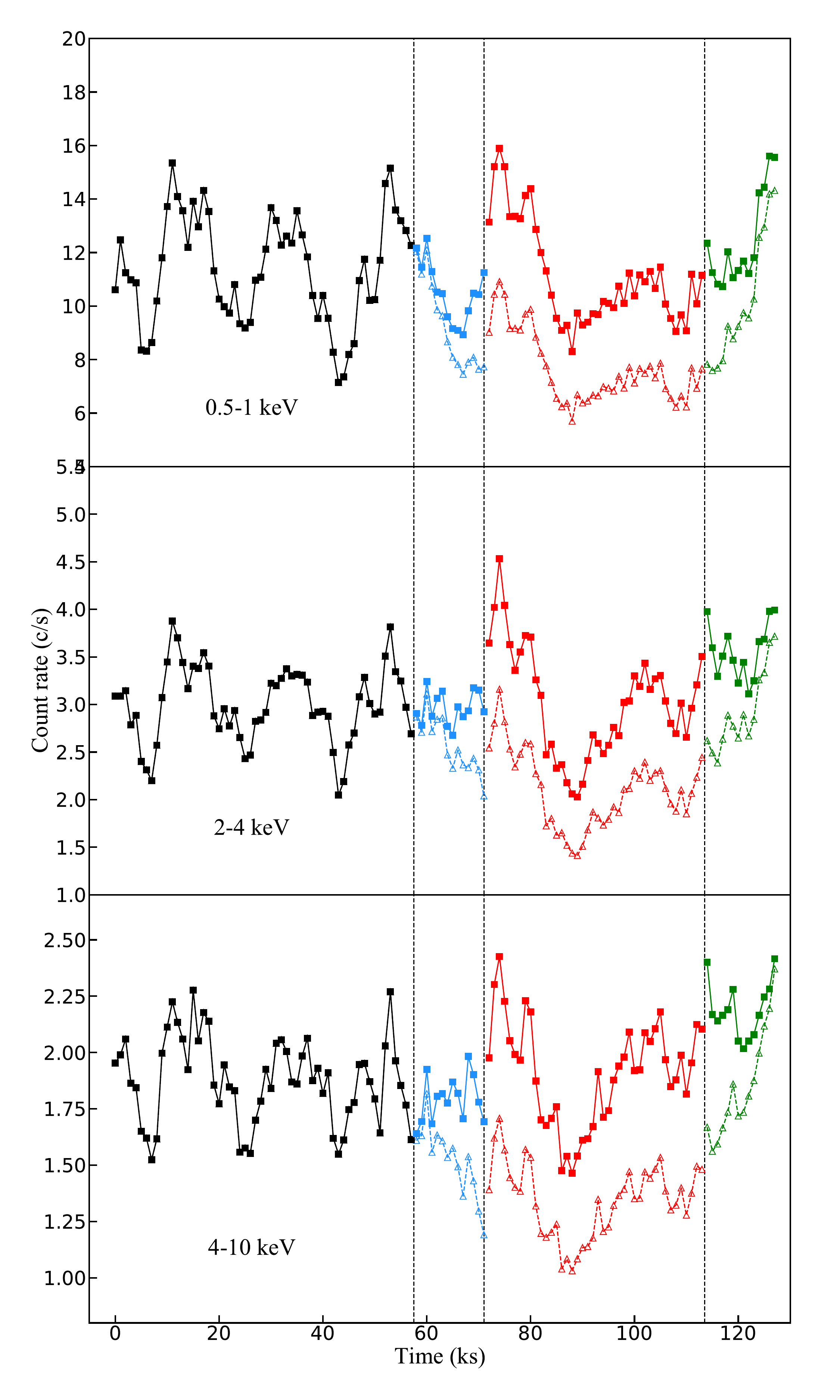}}
\caption{\label{fig:LC_corrected} The absorption corrected light curves of the eclipsed observation. Square/open triangle correspond to our final model with/without the Compton-thick component. Note for the $high$ period the triangles overlap with the squares. The absorption correction to the $ingress$ and $egress$ periods is time dependent through linear interpolating between the average correction factors of the $high$ and $low$ periods. }
\end{figure}  

\par Remarkably, the improved spectral fitting simultaneously reduces the deviation between periods/exposures in the absorption corrected HR -- CR plot (see now lower middle panel in Fig. \ref{fig:SR}), further supporting the validity of these revisions to the spectral fitting. Meanwhile, the occultation scenario is also manifested by the improved spectral fitting, with the covering fraction first increasing and then decreasing. However, two prominent features are still visible in the absorption corrected HR -- CR diagram: 1) the $low$ period lies along a somehow parallel track, though closer but still obviously offset (to the left) from that of the $high$ period; 2) the tracks of the $low$ and $high$ periods appear steeper than that of the $former$/$latter$ observations. 

\par Below we first tackle the offset between $low$ and $high$ periods within a single XMM exposure, which actually indicates two periods have similar spectral shape but the $low$ period is intrinsically fainter in flux. To further highlight this discrepancy, we plot the absorption-corrected light curves in Fig. \ref{fig:LC_corrected}. It is clear that the intrinsic (absorption corrected) count rate of the $low$ period is lower than that of the $high$ one, surprisingly by a similar factor of $\sim 30$ per cent in all three bands concerned. A possible but rather strained interpretation is that, the apparent $low$ period, anticipated to be caused by a severe eclipse, was coincidentally associated with an intrinsically low state that both the power law and the soft X-ray excess are dimmer by $\sim$ 30\%. This is already highly unlikely, not to mention the clear offset in the HR -- CR plot between $high$ and $low$ periods. 

\par Nevertheless, even if ignoring the constraint from the spectral fitting, the mere HR -- CR diagram leaves the model little room to adjust. During the spectral fitting, forcing the Compton-thin transient absorption to be stronger (i.e., with larger $N_{\rm H}$ and $f_{\rm cov}$ or smaller $\xi$) could shift the $low$ period to the right in the absorption-corrected HR -- CR plot, but at the same time shift it upwards since an intrinsically softer spectrum will be yielded, thus the offset between the $low$ and $high$ periods would remain unsolved. 

\par In this case, highly Compton-thick absorption which could achieve nearly full extinction all over the concerned energy band would be indispensable. Including such an additional partially covering Compton-thick absorber could easily raise the intrinsic flux level, without spoiling the spectral fitting or altering the fitting results (particularly the photon index and thus the HR) except for the normalization. In fact, transient Compton-thick absorbers have been reported in several sources \citep[e.g.][]{Risaliti_2007, Risaliti_2009, Turner_2011, Miniutti_2014}, and the concept of a multi-phase eclipsing cloud composed of both Compton-thin and Compton-thick components was already proposed in \citet{Risaliti_2011}. 

\par Assuming the Compton-thick absorber is so thick that it leaves no spectral imprint in the concerned energy band (0.5 -- 10 keV), in our final model we manually assign a covering factor of the Compton-thick absorber (with $N_{\rm H}$ = 5 $\times$ 10$^{24}$ cm$^{-2}$) for each period (0\%, 15\%, 30\%, and 15\% for the $high$, $ingress$, $low$ and $egress$ period respectively), presuming that no Compton-thick absorption exists during the $high$ period, and the average intrinsic flux level is the same in all the periods in the 2016 exposure. With this model, we could ease the offset between $low$ and $high$ periods in the HR -- CR plot (see the lower right panel of Fig. \ref{fig:SR}, and Table \ref{tab:results} for the parameters). Note we can not constrain such Compton-thick absorption through fitting the EPIC-pn spectra due to the limited energy band, and unfortunately no coordinated NuSTAR observation is available during the 2016 exposure. Manually adding the partial covering Compton-thick absorption with covering factor fixed at 30\% to the EPIC-pn spectral fitting would yield a lower limit to the column density $N_{\rm H}$. In this case, we derive a lower limit of $2.5 \times 10^{24}\, \rm cm^{-2}$ at 99\% confidence level ($\Delta \chi^2 \sim 7$). 

\par The last concerned feature in the absorption corrected HR -- CR plot (the lower right panel of Fig. \ref{fig:SR}) is that the ``softer-when-brighter" tracks of the $high$ and $low$ periods are steeper than that seen in the $former$ and $latter$ exposures. Does it mean the intrinsic spectral variation during the 2016 exposure behaves differently? 
As we will show in \S\ref{sec:discussion}, this feature could instead be a natural consequence of the transient absorption. 

\par
\begin{figure*}
\centering
\subfloat{\includegraphics[width=0.33\textwidth]{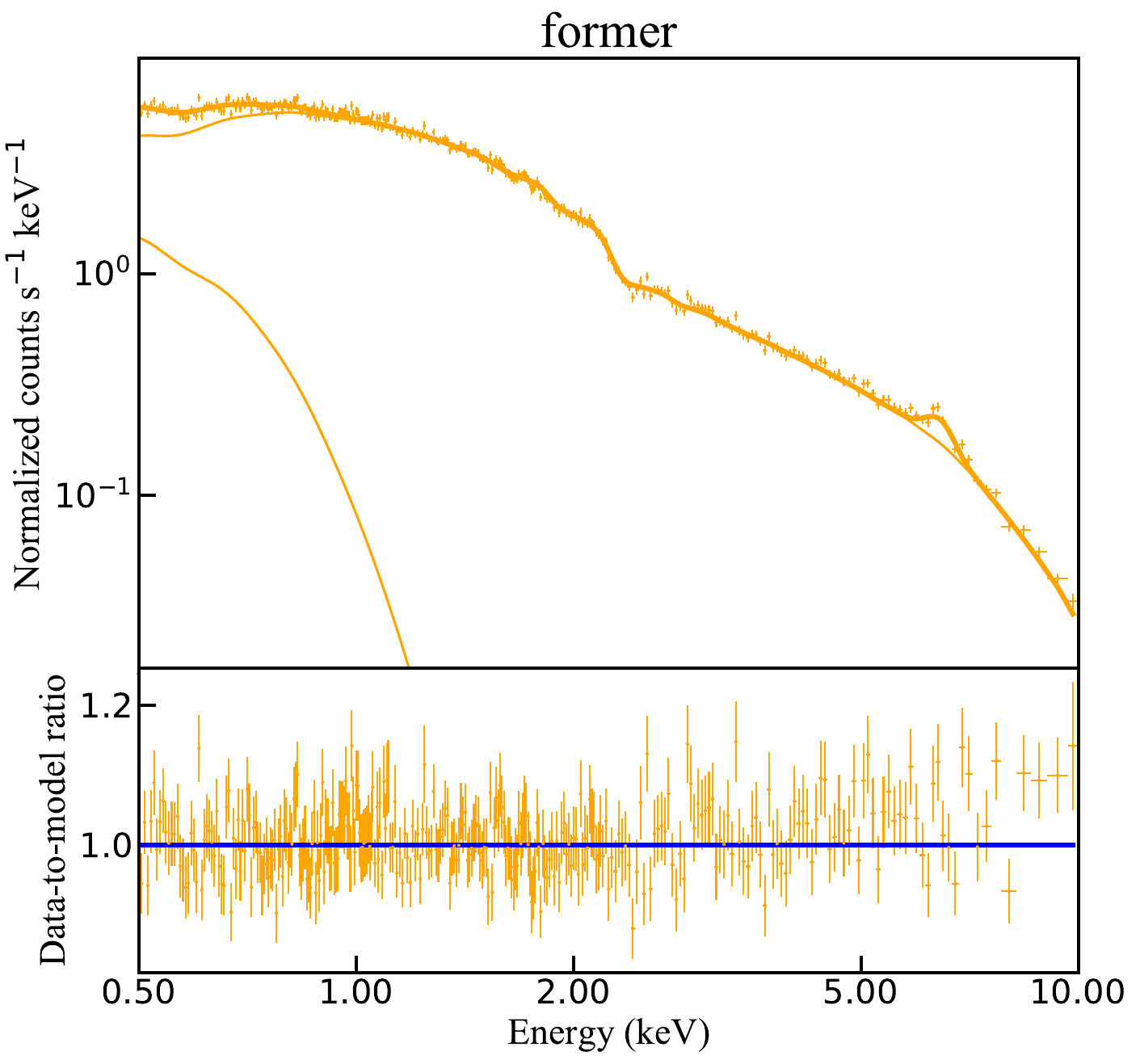}}
\subfloat{\includegraphics[width=0.33\textwidth]{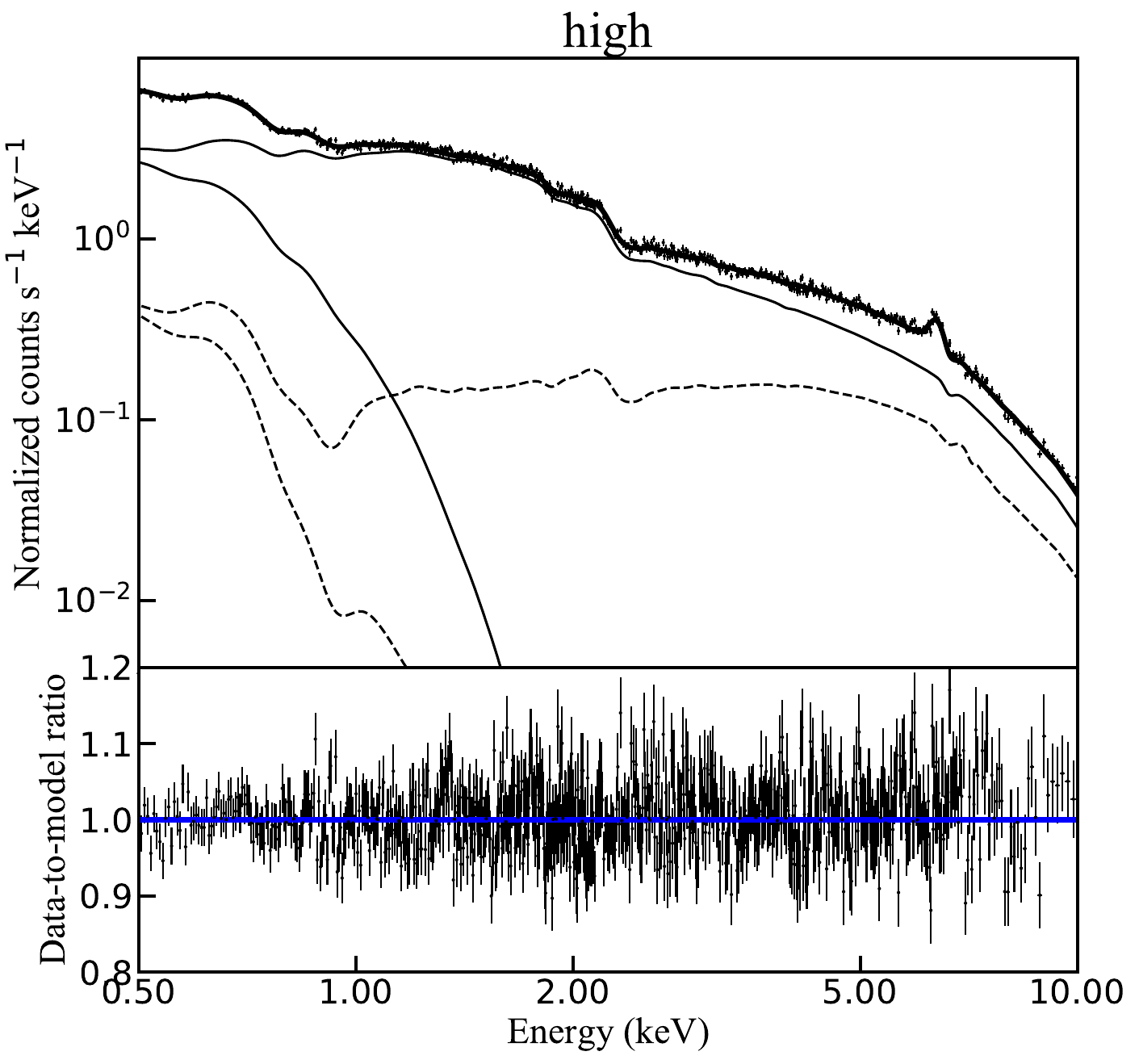}}
\subfloat{\includegraphics[width=0.33\textwidth]{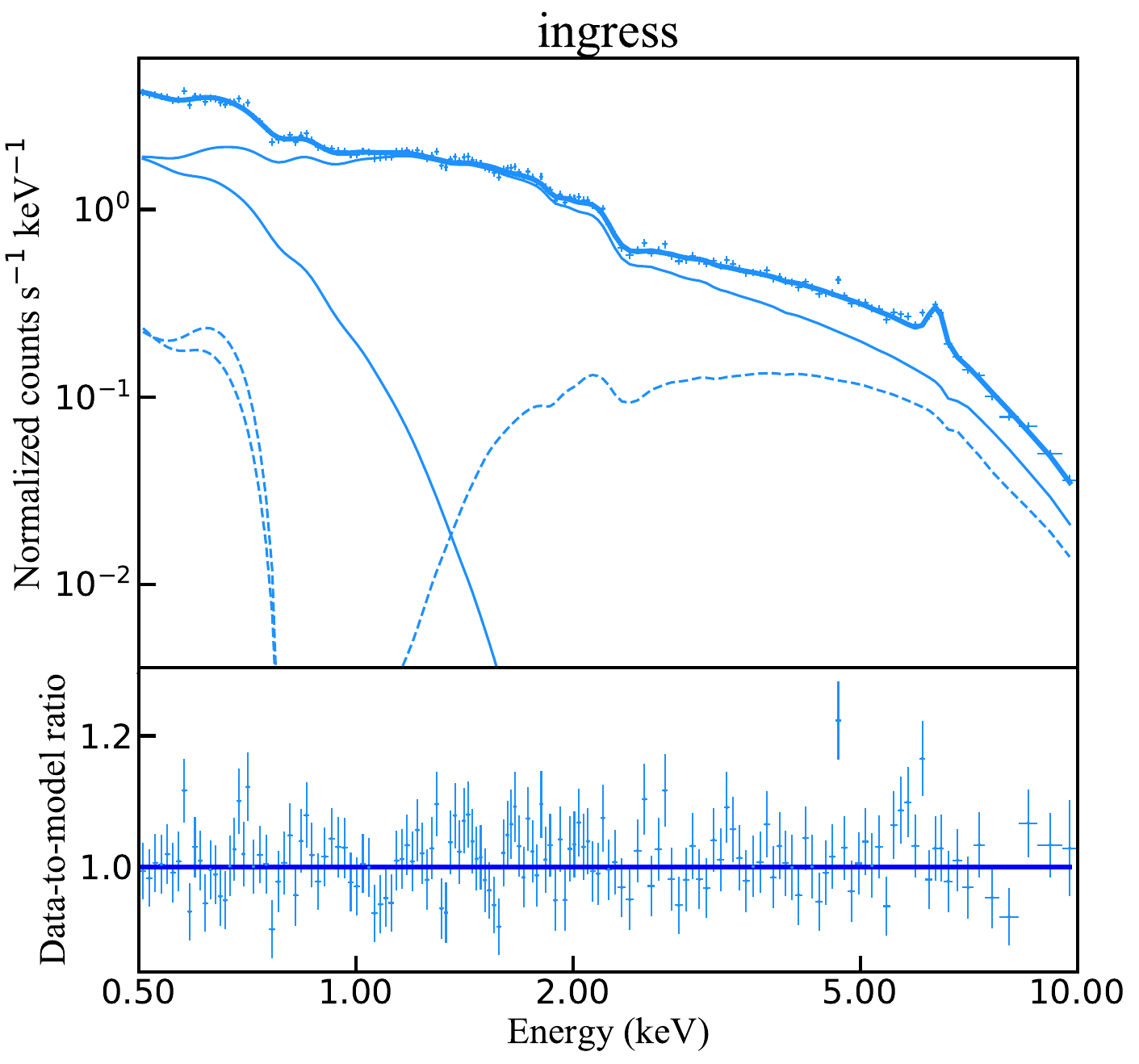}}\\
\subfloat{\includegraphics[width=0.33\textwidth]{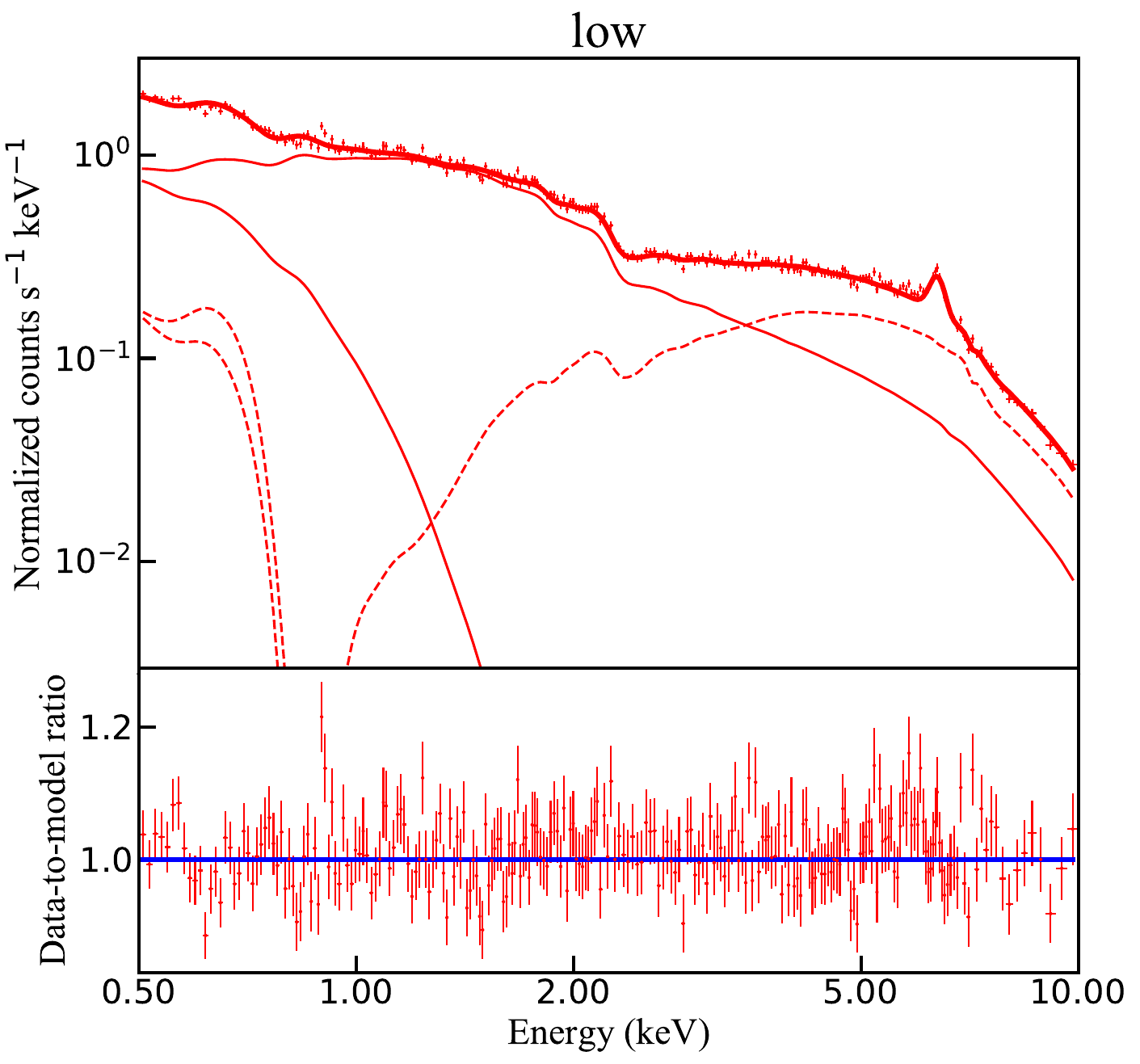}}
\subfloat{\includegraphics[width=0.33\textwidth]{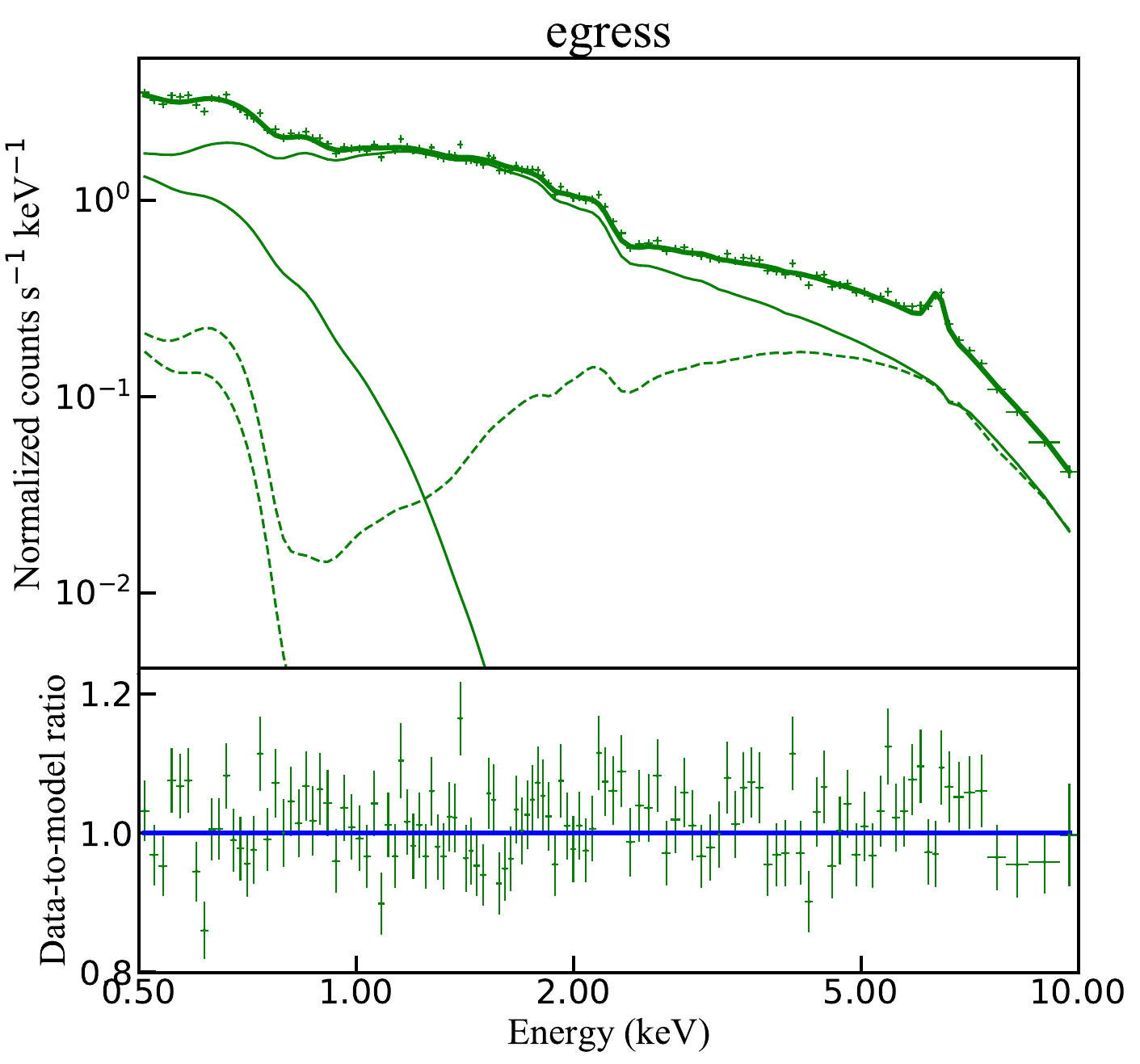}}
\subfloat{\includegraphics[width=0.33\textwidth]{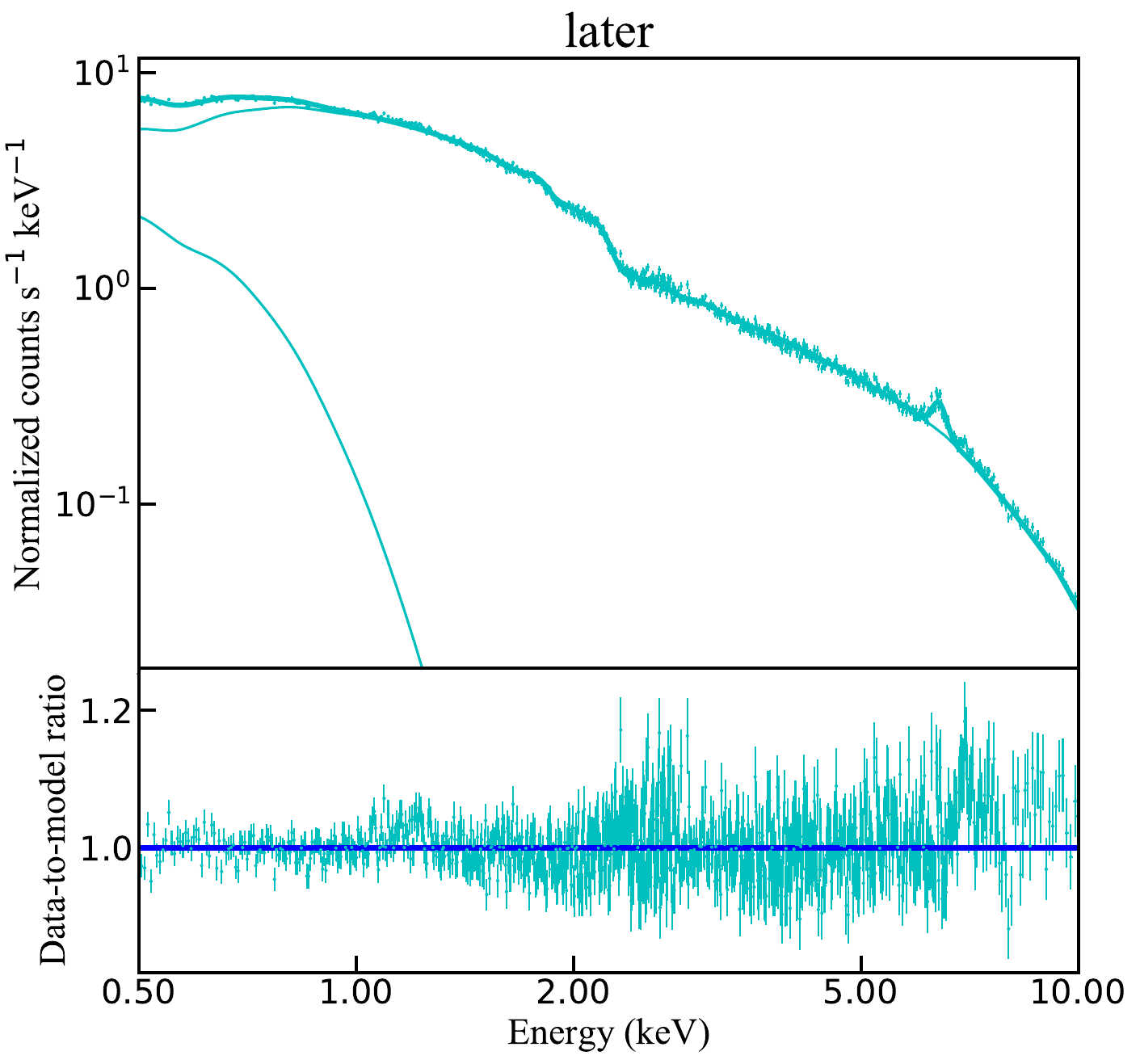}}
\caption{\label{fig:pn_spectra} The EPIC-pn spectra and the data-to-model ratios. The models of the soft X-ray excess and the power law continuum are plotted. Dotted/solid lines signify the emission absorbed/unabsorbed by the partially covering Compton-thin eclipsing absorber, while the bold line denotes the total model. Spectra are re-binned for better illustration.}
\end{figure*}

 Therefore, we conclude this final spectral model most appropriately describes this occultation event, and the corresponding fitting results are adopted to derive the size/geometry of the transient absorber in \S\ref{sec:discussion}. This best-fit model along with the EPIC-pn spectra is shown in Fig. \ref{fig:pn_spectra}. Finally, we also re-fit the RGS spectra through including the eclipsing absorber with parameters fixed to those we obtained from EPIC-pn spectra. The fitting is only slightly worse ($\Delta C \sim 20$ for the $high$ and $low$ spectra), and the best-fit parameters of the warm absorbers and the continuum are barely altered. In this sense, we obtain consistent fitting results between the RGS and EPIC-pn spectra. 

\section{Discussion}
\label{sec:discussion}

\par Following \citet{Reeves_2018, Turner_2018, Gallo_2021}, we constrain the geometry of the eclipsing system based on the spectral fitting results, under certain assumptions:\\

(i) $D = v \,\Delta t$ where $v$ is the velocity of the eclipsing absorber, $\Delta t = 56 $ ks is the duration of the occultation event, and D the size (diameter, assuming round shapes) of either the absorber or the corona (see further below). 

(ii) $v^2 = k\frac{GM}{r}$ where $M$ is the mass of the central black hole, $r$ the distance between the absorber and the black hole, and $k$ a scaling factor. $k = 1$ indicates a Keplerian orbit. For NGC 6418, the black hole mass $M_{\rm BH}$ is estimated to be $1.09 \times 10^7 M_{\odot}$ \citep{Bentz_2015}.

(iii) $n_{\rm H} = \frac{N_{\rm H}}{\alpha D}$ where $n_{\rm H}$ is the number density, $N_{\rm H}$ the average column density of the absorber, and $\alpha$ a scaling factor. $\alpha$ depends on the density distribution and the geometry of the cloud. Under a uniform density distribution, $\alpha$ is 1 for a cube, and 2/3 for a sphere. For complicated conditions where the Compton-thin/thick components are mixed, $\alpha$ should be calculated case by case. 

(iv) $n_{\rm H}\,\xi = \frac{L_{\rm ion}}{r^2}$ where $L_{\rm ion}$ is the luminosity of the ionizing continuum. In this work we adopt the unabsorbed $L_{\rm 0.1-200\,keV}$ in the $high$ period as an estimation for $L_{\rm ion}$. Setting redshift at 0.0052 \citep{Meyer_2004}, we derive a $L_{\rm ion} \approx 7 \times 10^{42}\rm\,erg\,s^{-1}$ using \textit{clumin} within XSPEC. 

and simultaneous consideration of (i)–(iv) derives the estimation of the location $r$,
\begin{equation}\label{eq1}
r^{\frac{5}{2}} =k\alpha (GM)^{\frac{1}{2}}\frac{L\Delta t}{N_{\rm H}\xi} \tag{v}
\end{equation}

\subsection{Depicting the eclipsing absorber} 
\subsubsection{The clumpy absorber}
\begin{figure}
\centering
\subfloat{\includegraphics[width=0.45\textwidth]{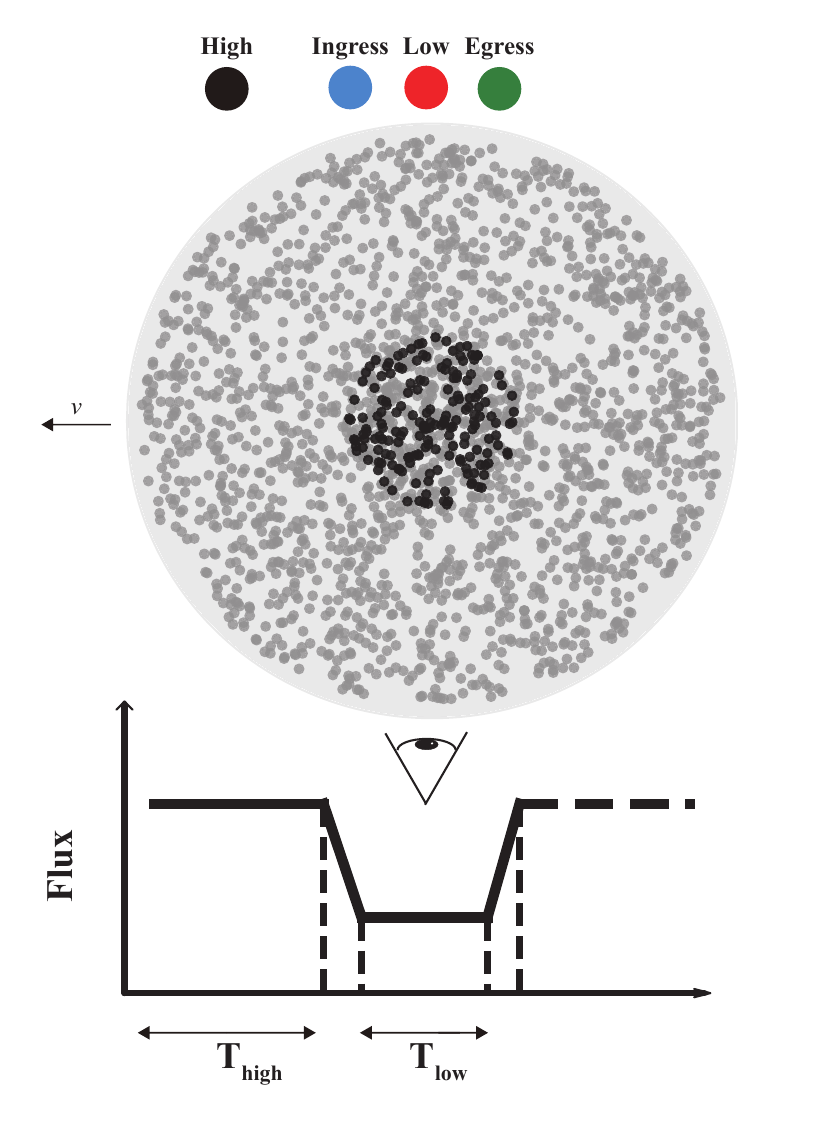}}
\caption{\label{fig:geometry} A schematic diagram showing the proposed scenario of the clumpy eclipsing system, composed of a dense core with both Compton-thin (grey dots) and Compton-thick (black dots) clouds, an outer sparser region with Compton-thin clouds, and two layers of warm absorbers (symbolized with light grey background). Note the whole system, the inner core and individual clouds are drawn to have round shapes just for simplicity, but not necessarily. The large black, blue, red, green dots on the top represent the corona in different states. The number, size and covering factor of the clouds, and the size of the corona are set to scale accordingly.}
\end{figure}  

\par To apply these relations, we should first determine the geometry of the eclipsing system. \citetalias{Gallo_2021} proposes the simplest condition that a single homogeneous cloud partially eclipses the corona (see Figure 1 of \citetalias{Gallo_2021}; in this case $v \,\Delta t$ measures the diameter of the corona). In \citetalias{Gallo_2021}, the corona size is derived as $\frac{\rm D_{corona}}{\rm D_{absorber}} = \frac{\Delta t_{\rm low}}{\Delta t_{\rm ingress}} = \frac{\rm 43\,ks}{\rm 13\,ks} = 3.3$, which means the eclipsing cloud can block at most 10\% (1/3.3$^2$) of the corona, seriously contradicting the high covering factor of the transient absorber yielded from spectral fitting. Note that, strictly, in this scenario the equation should be $\frac{\rm D_{corona}}{\rm D_{absorber}} = \frac{\Delta t_{\rm low}+\Delta t_{\rm ingress}}{\Delta t_{\rm ingress}}$ = 4.3, where $\Delta t_{\rm low}+\Delta t_{\rm ingress}$ is defined as the duration of the eclipse $\Delta t$, making the issue even worse.

\par We point out that, for an occultation event, $\frac{\Delta t_{\rm low}+\Delta t_{\rm ingress}}{\Delta t_{\rm ingress}}$ either equals $\frac{\rm D_{corona}}{\rm D_{absorber}}$, or its reciprocal$\frac{\rm D_{absorber}}{\rm D_{corona}}$. In the later case which shall apply here (see Fig. \ref{fig:geometry}), the absorber has size 4.3 $\times$ that of the corona (but not the other way around), and $v\Delta t$ derives the diameter of the absorber (instead of the corona). Moreover, the absorber has to be clumpy and composed of smaller clouds (but see Appendix for alternative but improbable geometry settings) to stay accord with the partially eclipsing scenario. 

\par The next step is to delineate such a cluster of clouds with the spectral fitting results. 
We first focus on the eclipse event defined with the $ingress$, $low$ and $egress$ periods. As we have shown above, the absorber has to be clumpy, with both Compton-thin and Compton-thick absorption. We plot in Fig. \ref{fig:geometry} a diagram for such an absorber (the central denser region in the plot). The covering factor of the Compton-thick absorption during the $low$ period is 30\%, and that of the Compton-thin absorber is (1-30\%)*0.76 = 53\%, where 0.76 is the covering factor derived from spectral fitting neglecting the Compton-thick absorption (namely the $f_{\rm cov}$ of zxipcf in the bottom half of Table \ref{tab:results}).

\par Unfortunately, unlike a single homogeneous cloud, the complexity of clumpy absorber hinders the determination of its properties. A pivotal parameter, the number of those discrete clouds, or the radius of each cloud, is lacking. This parameter is essential for the calculation of the $\alpha$ in Equation (\ref{eq1}), which relates the column density $N_{\rm H}$ to the volume density $n_{\rm H}$. In principle, the radius of each discrete cloud should not be too large, otherwise we could see the ingress/eclipse/egress of each individual cloud thus contradict the smooth ingress/egress light curves we observed (see Fig. \ref{fig:LC}); but also not too small, otherwise the $n_{\rm H}$ has to be abnormally large to produce the observed $N_{\rm H}$. However, quantitative determination is infeasible. We show a typical case in Fig. \ref{fig:geometry}, where the individual clouds have radii one-twentieth of that of the inner cluster. The surface density of the Compton-thin and Compton-thick clouds is then chosen to reproduce the observed covering factors. In case of no overlapping, there would be $\sim 210$ Compton-thin and 120 Compton-thick clouds in the inner cluster. 

\par In this case, the parameter $\alpha$ is calculated to be 0.033. Supposing a Keplerian orbit (k=1), Equation (\ref{eq1}) places the eclipsing clouds at $r \approx 1.0 \times 10^{10}\,km = 641\,r_{\rm g}$, where $r_{\rm g}$ is defined as $\frac{GM}{c^2}$. We can further derive other properties using Equation (i)–(iv): $D_{\rm inner\,cluster} \approx 40\,r_{\rm g}$, $v \approx 1.2 \times 10^{4}\,km\,s^{-1}$, $D_{\rm corona} \approx 9.5\,r_{\rm g}$, and $n_{\rm H} \approx 7.8 \times 10^{10}\,cm^{-3}$ (of the Compton-thin cloud, hereafter the same\footnote{Assuming the Compton-thick cloud has the same size as the Compton-thin clouds, its density shall be larger by a factor of $>$ 14.5 (the ratio of the column densities of Compton-thick and thin clouds).}). Compared with that in \citetalias{Gallo_2021}, the individual cloud here is smaller, but with larger column density and ionization parameter, which indicates a larger volume density and a shorter distance to the ionization source. For comparison, the $\sigma_{\rm line}$ of the broad H$\beta$ of NGC 6814 is $1918 \pm 36 \,km\,s^{-1}$ \citep{Bentz_2009}. Therefore, the absorber is likely located in the high-ionization broad line region (BLR) or further inner \citep[see Figure 1.1 in][]{Gallo_2023}. 

\par Meanwhile, although the scenario here is quite different from \citetalias{Gallo_2021}, we derive a similarly small corona with $D_{\rm corona} \approx 9.5 \,r_{\rm g}$. This value agrees well with those reported in other sources using independent methods \citep[e.g., ][]{Risaliti_2009, Parker_2014,Wilkins_2015, Gallo_2015, Caballero_2018, Alston_2020, Hancock_2023}. Though this picture seems reasonable, we note that large uncertainty exists in such an analysis. The estimation utilizes the parameters of the spectral fitting, which are model dependent and could be degenerated with each other. Moreover, the division of the four periods is somewhat arbitrary, which directly influences the derived coronal size. Furthermore, here we assume $D_{\rm individual~cloud}/D_{\rm inner~cluster}= 1/20$. A larger value would result in a larger $r$, smaller $n_{\rm H}$, $D_{\rm inner cluster}$ and $D_{\rm corona}$, and vice versa. However, as long as we stick to the scenario where a large clumpy cluster eclipses the corona, the estimated $D_{\rm corona}$ is always small even considering these uncertainties. For example, a diameter ratio of 1/4 or 1/100 will result in $D_{\rm corona} \approx 7$ or $13\,r_{\rm g}$, $r \approx 1200$ or $340\,r_{\rm g}$, $v \approx 0.8 $ or $  1.6 \times 10^{4}\,km\,s^{-1}$, $n_{\rm H} \approx 2 $ or $ 28 \times 10^{10}\,cm^{-3}$, respectively. 

\par We further note that partially covering absorption is also statistically detected in the $high$ period, with comparable $N_{\rm H}$ but considerably lower covering factor than the $low$ period. This means at the beginning of the 2016 exposure, the corona has already been eclipsed. This is further supported by the fact that, in the HR -- CR plot, the $high$ period would severely deviate from the $former$/$latter$ exposures before correcting the partially covering absorption (upper left panel of Fig. \ref{fig:SR}), and such deviation almost disappears after the absorption correction (see also \S\ref{sec:4.2}).  Such absorption to the $high$ period, with $N_{\rm H}$ comparable to that of the eclipsed periods, is also transient since it is not detected in either $former$ or $latter$ exposures. It also has to be clumpy considering the long duration of the high period (58 ks). Therefore it could naturally be the outer and sparser region of the eclipsing absorber we discussed above (see Fig. \ref{fig:geometry}). Unfortunately, we can only give a lower limit to its diameter ($>$ 3.5 $\times$ of the denser core) as we have not viewed its ingress/egress.  

\subsubsection{The warm absorbers}\label{sec:4.1.2}

\par Finally, we try to incorporate the two warm absorbers identified in the RGS spectra into this system. We find the warm absorbers (especially $xabs2$) become weaker during the $low$ period and are absent in the $former$ and $latter$ observations, indicating the warm absorbers are also transient, likely physically associated with the transient clumpy absorber we identified. The outflowing velocities of the two warm absorbers are also comparable to the Keplerian velocity of the transient clumpy absorber. 

\par
\begin{figure}
\centering
\subfloat{\includegraphics[width=0.45\textwidth]{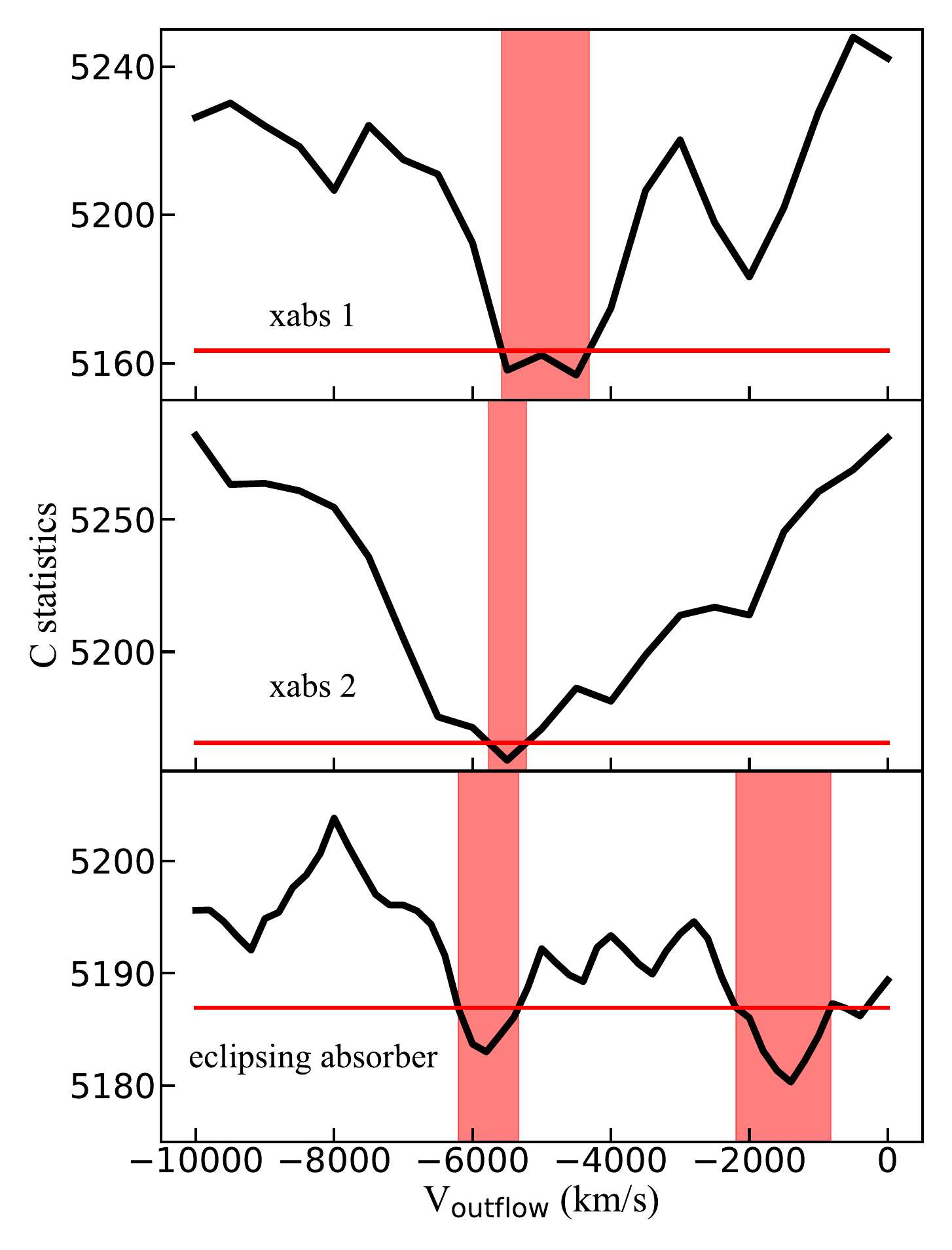}}
\caption{\label{fig:vout} The confidence regions of the outflow velocity of the two warm absorbers $xabs1$ and $xabs2$, and the eclipsing absorber, derived by fitting the RGS spectra. The red shaded areas indicate the 99\% confidence ranges (with $\Delta C = 6.6$).    
}
\end{figure}

 Therefore, we make an effort to measure the outflow velocity of the eclipsing clumpy absorber to investigate its potential connection with the warm absorbers. Unfortunately the EPIC-pn spectra have too poor spectral resolution to constrain the outflow velocity of the transient clumpy absorber. We revisit the RGS spectra through adding a third but partially covering $xabs$ component to model the effect of the transient clumpy absorber. 
But note the emission transmitted through the clumpy absorber (see the dashed line in Fig. \ref{fig:pn_spectra}) is rather weak within the RGS bandwidth, and thus this third $xabs$ could be statistically non-required in RGS spectra.
We hence only consider the $high$ interval during which the clumpy absorber has lower column density so that the transmitted X-ray emission may have notable contribution to the total spectra within the RGS bandwidth (see Fig. \ref{fig:pn_spectra}), and enforce the column density, ionization parameter and covering factor of the third $xabs$ absorber to vary only within the 90\% confidence ranges derived from the EPIC-pn spectra. As shown in Fig. \ref{fig:vout}, the outflowing velocity of this enforced absorber, though not as well constrained as the two warm absorbers, shows two local optimum ($\sim$ 5800 and 1600 $km\, s^{-1}$). Remarkably, the outflow velocities of the three absorbers are statistically consistent within 99\% confidence ranges. 

\par
\begin{figure}
\centering
\subfloat{\includegraphics[width=0.45\textwidth]{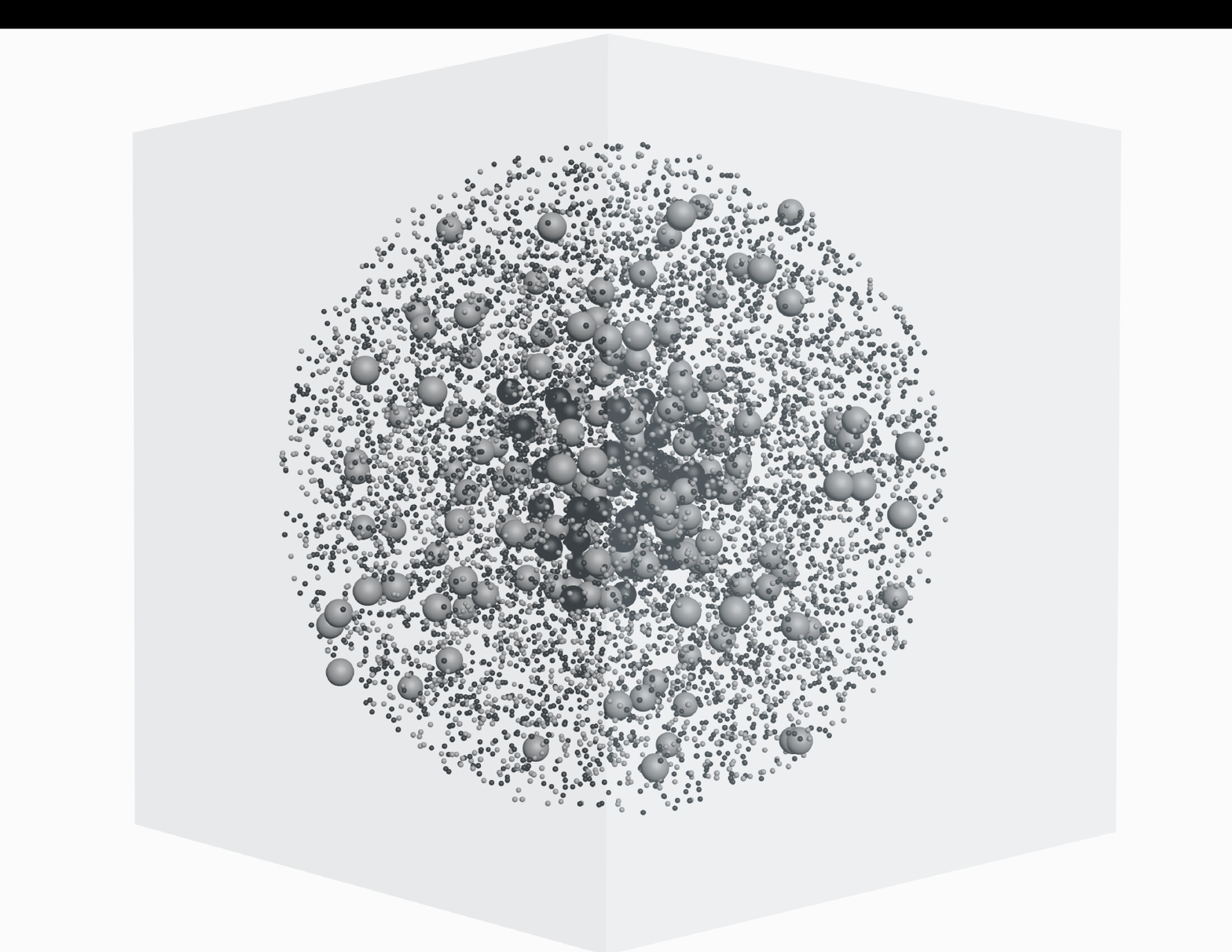}}
\caption{\label{fig:3Dcloud} A three-dimensional cartoon of the marvellous absorber, composed of Compton-thin/Compton-thick clouds (larger grey/black spheres), and high/low ionized warm absorbers (smaller grey/black spheres).  Note again the whole system, the inner core and individual clouds are drawn to have round shapes just for simplicity, but not necessarily. The cartoon is drawn for visualization, but not to scale (including the number, size and covering factor of the clouds). A three-dimensional video is also provided in the online material. 
}
\end{figure}

 Therefore, we infer the three absorbers are physically associated. Supposing they are located at the same distance, and ignoring their obscuration effect to each other, the density of the warm absorbers ($n_{\rm H} = \frac{L_{\rm ion}}{r^2\xi}$) would be $\sim$ 10 times that of the transient clumpy Compton-thin clouds for $xabs1$, while $\sim$ 0.1 times for $xabs2$. In this case, the significantly smaller $N_{\rm H}$ of the warm absorbers indicate small size of individual warm absorber clouds, i.e., can not exceed $\sim$ 1/400 and 1/2 (for $xabs1$ and $xabs2$ respectively) of the $D_{\rm individual~cloud}$. The two warm absorbers thus are also clumpy\footnote{ We try to constrain the covering factors of the warm absorbers. For $xabs2$, both the RGS and EPIC-pn spectra statistically support a full-covering absorption ($f_{\rm cov} > 0.80$, and $>$ 0.95, respectively). For $xabs1$, we derive a $f_{\rm cov}$ of $0.77_{-0.02}^{+0.06}$ when fitting the RGS spectra, while the EPIC-pn spectra still support a full-covering absorption ($f_{\rm cov} > 0.85$). Such a discrepancy could likely be attributed to the oversimplified continuum model when fitting the RGS spectra.}. They could be ablated, or tidal stretched/disrupted fragments from larger clouds, like the ``comet dust'' around comets. The low ionized ($xabs1$) and high ionized ($xabs2$) warm absorbers could initially originate from the Compton-thick (higher density) and Compton-thin (lower density) clouds, respectively. The much smaller column density of $xabs1$ suggests ablation/stretching/disruption is harder for the higher density clouds. The ablation/stretching/disruption could also be weaker in the denser core, thus yielding lower column densities of the warm absorbers during the $low$ interval. A three-dimensional model of this composite absorber is shown in Fig. \ref{fig:3Dcloud} and a video in the online material. 

\par Here we briefly compare the two ``warm absorbers" with the canonical warm absorbers reported in literature. Generally, the classic warm absorbers have a typical outflow velocity of $100 - 6000\,km\,s^{-1}$, a distance of $0.01 - 400\,pc$ and a number density of $10^{3} - 10^{8}\,cm^{-3}$ \citep[e.g., ][]{Krongold_2007, Steenbrugge_2009, Kaastra_2012, Longinotti_2013, Laha_2014, Ebrero_2021, Wang_2022}. Meanwhile, long-term monitoring of several sources shows these warm absorbers do not appear/disappear over time \citep[e.g.,][]{Krongold_2010, Silva_2016, Mao_2019}. For comparison, the two ``warm absorbers" in this work have a outflow velocity of $\sim 5000\,km\,s^{-1}$, a distance of 0.4 ld or $4 \times 10^{-4}$ pc and a number density of $10^{10}-10^{12}\,cm^{-3}$ (for $xabs2$ and $xabs1$ respectively), and appear/disappear among observations. Therefore, the two ionized absorbers in this work are likely not the traditional warm absorbers; instead, their properties are more similar to the so-called obscuring wind \citep{Kaastra_2014, Mehdipour_2017}. Notably, this transient absorber system is similar to that of the patchy transient obscurer reported in NGC 5548 \citep{Kaastra_2014, DiGesu_2015} which is also composed of a mixture of ionized gas with embedded colder and denser parts, and likely located close to the inner BLR. 

\subsection{An absorption-variation dominated ``softer-when-brighter" trend}\label{sec:4.2}

\begin{figure}
\centering
\subfloat{\includegraphics[width=0.5\textwidth]{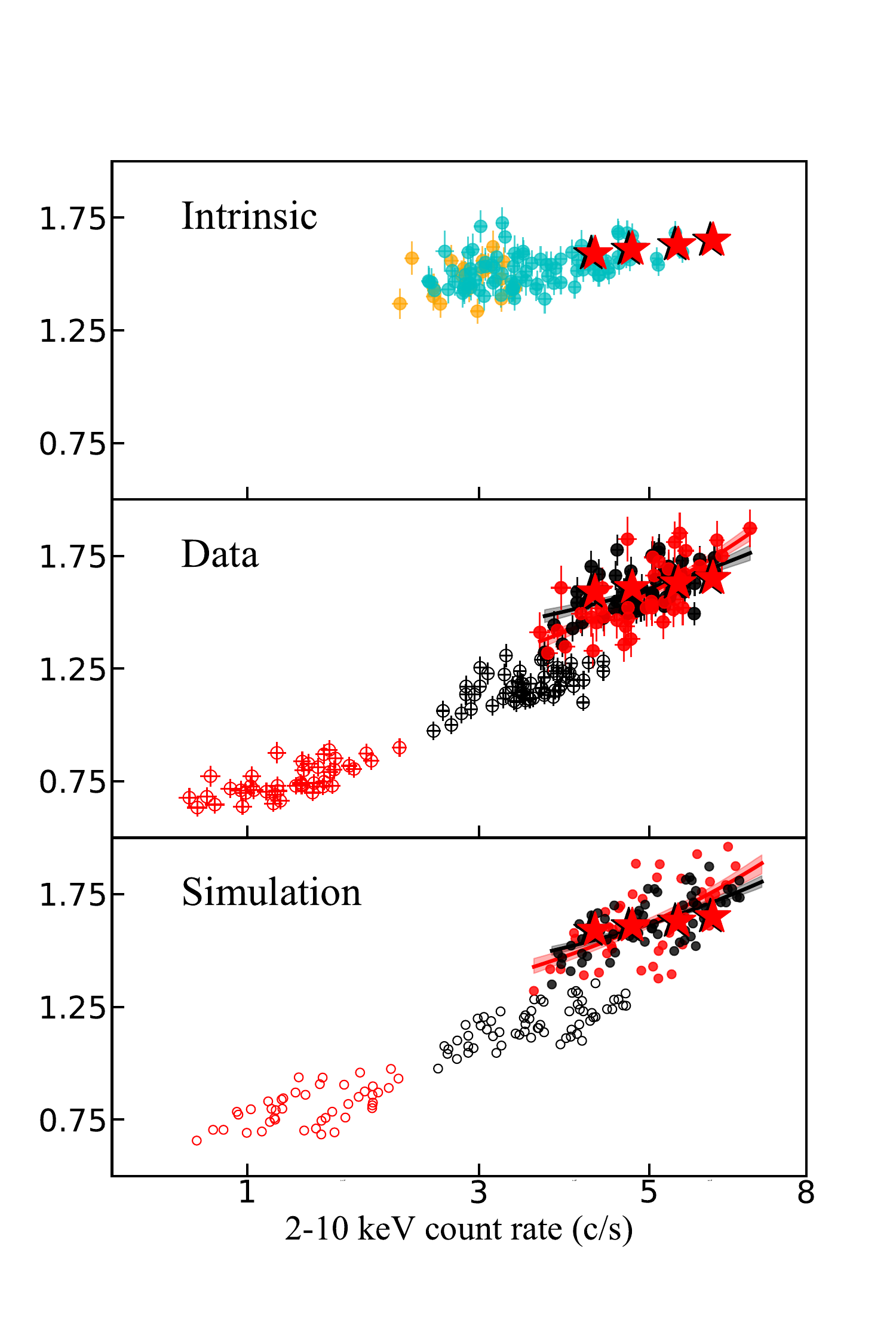}}
\caption{\label{fig:simulation} Upper panel: we adopt four typical values (marked as red/black stars) as the presumed intrinsic HR -- CR of the $low$/$high$ periods used as input to simulations. The four points are chosen to follow the track of the $former$ and $latter$ exposures (orange and cyan). Middle panel: an integration of the HR -- CR diagrams in Fig. \ref{fig:SR}, presented here for direct comparison. The observed/absorption-corrected data for the $low$ and $high$ intervals (red and black) are shown with open/solid circles. Lower panel: a simulation to show that the transient absorption could well reproduce the observed  artificial ``softer-when-brighter" pattern. Correcting the average absorption to each period would yield steeper HR -- CR tracks than input, which are caused by subtle variation of absorption within each period. Solid lines are simple linear regression to the absorption-corrected HR -- CR from the data and simulations, to highlight the tracks.
}
\end{figure}  

\par Finally we return to the last concerned feature in the absorption corrected HR -- CR plot we have not yet interpreted: the ``softer-when-brighter'' tracks of the $high$ and $low$ periods are steeper than that of the $former$ and $latter$ exposures.
We show below that this could naturally be interpreted as the result of subtle to moderate variation of the absorption within each period. In the time-resolved spectroscopy, by dividing the time sequence into four periods, we could only acquire the average absorption in each period (thus average correction for the absorption). However, as we now have realized that the clumpy eclipsing absorption (not merely the covering factor but also the column density) varies between the periods, the absorption could also subtly vary within each period, particularly considering the clumpy nature of the absorber. 

\par Note the diagram shown in Fig. \ref{fig:geometry} could commendably account for the varying absorption. Although we use the same marker for all the Compton-thin clouds for simplicity, the property can differ from cloud to cloud, which could naturally reproduce the slightly different best-fit column densities for different periods. Meanwhile, the random distribution of the clouds could also easily cause moderate variation of the absorption's covering factor within each period (see Fig. \ref{fig:simulation}). 

\par We then perform Monte Carlo simulation to illustrate this effect. We assume the intrinsic HR -- CR shapes of the $high$ and $low$ periods are similar to those of the $former$ and $latter$ observations, i.e., rather flat in the HR -- CR diagram. Starting from four typical data points (stars in Fig. \ref{fig:simulation}), we simulate a bunch of `observed' data absorbed by absorption with parameters randomly varying around the best-fit values, add Poisson fluctuations, and finally apply the average absorption correction. For the simulations plotted in Fig. \ref{fig:simulation}, the variation ranges of ($N_{\rm H}$, log $\xi$,  $f_{\rm cov}$) are ([11, 13], [1.9, 2.1], [0.3, 0.5]) and ([16, 18], [1.8, 2.0], [0.7, 0.85]) for the $high$ and $low$ periods, respectively. Commendably, we find that the simulated HR -- CR tracks, both `observed' and `absorption-corrected',  could well reproduce those of the data. After `absorption-correction', the simulated tracks of the $high$ and $low$ periods do appear steeper than the input shape, just as revealed by the data. 

\par It is thus reasonable to believe that, during the eclipsed exposure, the observed ``softer-when-brighter" trend is dominated by the variation of the absorption, and the intrinsic flux variability of the source during the exposure is much weaker than directly observed. The facts that the absorption-corrected count rates are now similar in all the four periods within all energy bands (see Fig. \ref{fig:LC_corrected}), and that the intrinsic fluxes of both the soft excess and the power law continuum are similar in all the periods (see Table \ref{tab:results}), further support the scenario of weak intrinsic variability. We finally note that, in case that the flux variability during one exposure is dominated by the absorption variation, the yielded artificial ``softer-when-brighter" trend could appear ``normal". It would then be hard to identify the occultation event based on the HR -- CR plot from one exposure alone. As illustrated in this work, in additional to light curves and spectral fitting, HR -- CR diagrams from more exposures for the same source could significantly help.  

\appendix

\section{Alternative but improbable geometric settings}

\par Here we discuss alternative but unlikely geometric settings for the eclipse absorber, assuming the absorber is a single cloud but not clumpy.  

\par
\begin{figure}
\centering
\subfloat{\includegraphics[width=0.45\textwidth]{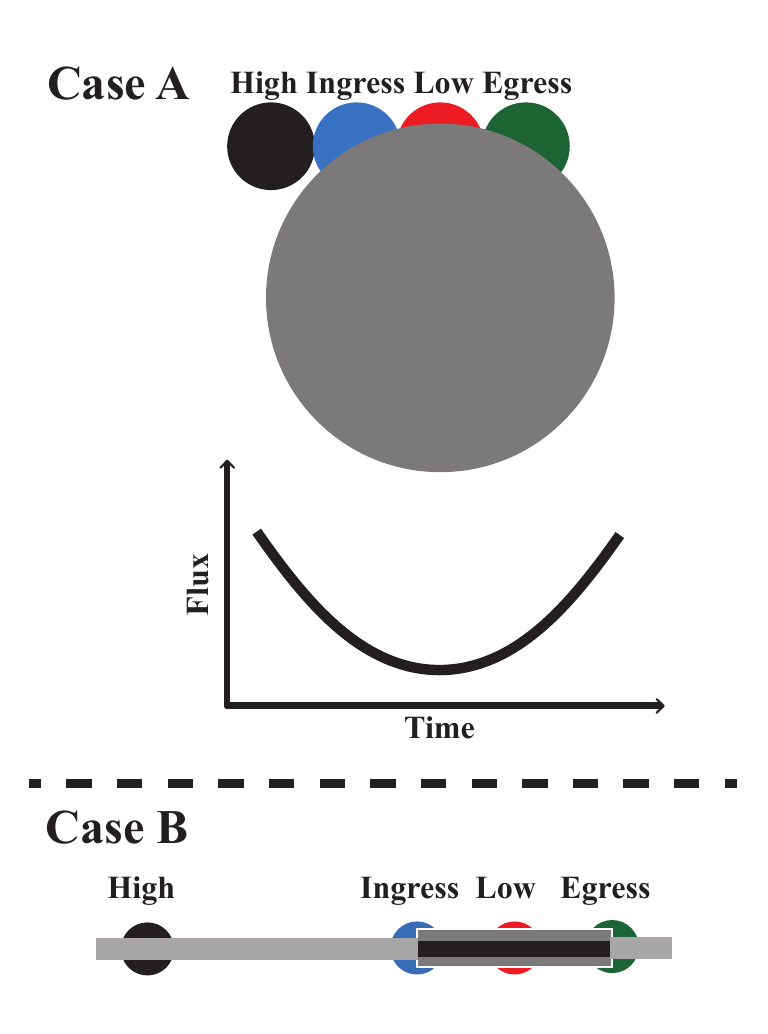}}
\caption{\label{fig:alternative} Two alternative but unlikely geometric settings. A: a round absorber partially eclipsing the corona would yield rather smooth eclipse curve. The Compton-thick absorption is not even considered. B: a long and thin (``comet''-like) absorber, with thinner ends, thicker cylinder-like central region with both Compton-thin (grey) and Compton-thick (black) absorption.}   
\end{figure}

 As shown in Fig. \ref{fig:alternative}, a partial eclipse by a single round cloud to a small corona may explain the observed large covering factor of the occultation event.  However, the eclipse curve would be rather smooth (unlike the observed sharp ingress and egress processes shown in Fig. \ref{fig:LC}). In this model. it is also difficult to incorporate the Compton-thick absorption, and the absorption with smaller covering factor during the $high$ period. 

\par If the absorber has a thin and long shape, such the ``comet"-like absorber reported in NGC 1365 \citep{Maiolino_2010}, the eclipse curve could be as sharp as the observed one. At least one end of the absorber needs to be thinner to explain the smaller covering factor observed during the $high$ period, and quickly thickens to reproduce the observed sharp ingress. The cylinder like central thicker region of the absorber should also have a Compton-thick inner cylinder. 
Putting all these requirements makes this geometric setting extremely unlikely. 

\section*{Acknowledgements} 
The work is supported by National Natural Science Foundation of China (grants No. 11890693, 12033006 $\&$ 12192221). The authors gratefully acknowledge the support of Cyrus Chun Ying Tang Foundations. The work is based on observations obtained with XMM-Newton, an ESA science mission with instruments and contributions directly funded by ESA Member States and NASA.

\section*{Data Availability}
The raw data used in this article are all public and available via the XMM-Newton Science Archive (\url{https://www.cosmos.esa.int/web/x mm-newton/xsa}). Other data underlying this article are available in the article and in its online supplementary material.

\bibliographystyle{mnras}
\bibliography{occultation}{}

\begin{thebibliography}{}
\makeatletter
\relax
\def\mn@urlcharsother{\let\do\@makeother \do\$\do\&\do\#\do\^\do\_\do\%\do\~}
\def\mn@doi{\begingroup\mn@urlcharsother \@ifnextchar [ {\mn@doi@}
  {\mn@doi@[]}}
\def\mn@doi@[#1]#2{\def\@tempa{#1}\ifx\@tempa\@empty \href
  {http://dx.doi.org/#2} {doi:#2}\else \href {http://dx.doi.org/#2} {#1}\fi
  \endgroup}
\def\mn@eprint#1#2{\mn@eprint@#1:#2::\@nil}
\def\mn@eprint@arXiv#1{\href {http://arxiv.org/abs/#1} {{\tt arXiv:#1}}}
\def\mn@eprint@dblp#1{\href {http://dblp.uni-trier.de/rec/bibtex/#1.xml}
  {dblp:#1}}
\def\mn@eprint@#1:#2:#3:#4\@nil{\def\@tempa {#1}\def\@tempb {#2}\def\@tempc
  {#3}\ifx \@tempc \@empty \let \@tempc \@tempb \let \@tempb \@tempa \fi \ifx
  \@tempb \@empty \def\@tempb {arXiv}\fi \@ifundefined
  {mn@eprint@\@tempb}{\@tempb:\@tempc}{\expandafter \expandafter \csname
  mn@eprint@\@tempb\endcsname \expandafter{\@tempc}}}

\bibitem[\protect\citeauthoryear{{Alston} et~al.,}{{Alston}
  et~al.}{2020}]{Alston_2020}
{Alston} W.~N.,  et~al., 2020, \mn@doi [Nature Astronomy]
  {10.1038/s41550-019-1002-x}, \href
  {https://ui.adsabs.harvard.edu/abs/2020NatAs...4..597A} {4, 597}

\bibitem[\protect\citeauthoryear{Anders \& Grevesse}{Anders \&
  Grevesse}{1989}]{ANDERS_1989}
Anders E.,  Grevesse N.,  1989, \mn@doi [Geochimica et Cosmochimica Acta]
  {10.1016/0016-7037(89)90286-X}, 53, 197

\bibitem[\protect\citeauthoryear{{Arnaud}}{{Arnaud}}{1996}]{Arnaud_1996}
{Arnaud} K.~A.,  1996, in {Jacoby} G.~H.,  {Barnes} J.,  eds,  Astronomical
  Society of the Pacific Conference Series Vol. 101, Astronomical Data Analysis
  Software and Systems V. p.~17

\bibitem[\protect\citeauthoryear{{Behar}, {Sako}  \& {Kahn}}{{Behar}
  et~al.}{2001}]{Behar_2001}
{Behar} E.,  {Sako} M.,   {Kahn} S.~M.,  2001, \mn@doi [\apj] {10.1086/323966},
  \href {https://ui.adsabs.harvard.edu/abs/2001ApJ...563..497B} {563, 497}

\bibitem[\protect\citeauthoryear{{Bentz} \& {Katz}}{{Bentz} \&
  {Katz}}{2015}]{Bentz_2015}
{Bentz} M.~C.,  {Katz} S.,  2015, \mn@doi [\pasp] {10.1086/679601}, \href
  {https://ui.adsabs.harvard.edu/abs/2015PASP..127...67B} {127, 67}

\bibitem[\protect\citeauthoryear{{Bentz} et~al.,}{{Bentz}
  et~al.}{2009}]{Bentz_2009}
{Bentz} M.~C.,  et~al., 2009, \mn@doi [\apj] {10.1088/0004-637X/705/1/199},
  \href {https://ui.adsabs.harvard.edu/abs/2009ApJ...705..199B} {705, 199}

\bibitem[\protect\citeauthoryear{{Bianchi}, {Piconcelli}, {Chiaberge},
  {Bail{\'o}n}, {Matt}  \& {Fiore}}{{Bianchi} et~al.}{2009}]{Bianchi_2009}
{Bianchi} S.,  {Piconcelli} E.,  {Chiaberge} M.,  {Bail{\'o}n} E.~J.,  {Matt}
  G.,   {Fiore} F.,  2009, \mn@doi [\apj] {10.1088/0004-637X/695/1/781}, \href
  {https://ui.adsabs.harvard.edu/abs/2009ApJ...695..781B} {695, 781}

\bibitem[\protect\citeauthoryear{{Caballero-Garc{\'\i}a}, {Papadakis},
  {Dov{\v{c}}iak}, {Bursa}, {Epitropakis}, {Karas}  \&
  {Svoboda}}{{Caballero-Garc{\'\i}a} et~al.}{2018}]{Caballero_2018}
{Caballero-Garc{\'\i}a} M.~D.,  {Papadakis} I.~E.,  {Dov{\v{c}}iak} M.,
  {Bursa} M.,  {Epitropakis} A.,  {Karas} V.,   {Svoboda} J.,  2018, \mn@doi
  [\mnras] {10.1093/mnras/sty1990}, \href
  {https://ui.adsabs.harvard.edu/abs/2018MNRAS.480.2650C} {480, 2650}

\bibitem[\protect\citeauthoryear{{Cash}}{{Cash}}{1979}]{Cash_1979}
{Cash} W.,  1979, \mn@doi [\apj] {10.1086/156922}, \href
  {https://ui.adsabs.harvard.edu/abs/1979ApJ...228..939C} {228, 939}

\bibitem[\protect\citeauthoryear{{Connolly}, {McHardy}, {Skipper}  \&
  {Emmanoulopoulos}}{{Connolly} et~al.}{2016}]{Connolly_2016}
{Connolly} S.~D.,  {McHardy} I.~M.,  {Skipper} C.~J.,   {Emmanoulopoulos} D.,
  2016, \mn@doi [\mnras] {10.1093/mnras/stw878}, \href
  {https://ui.adsabs.harvard.edu/abs/2016MNRAS.459.3963C} {459, 3963}

\bibitem[\protect\citeauthoryear{{Cox}, {Torres-Alba}, {Marchesi}, {Zhao},
  {Ajello}, {Pizzetti}  \& {Silver}}{{Cox} et~al.}{2023}]{Cox_2023}
{Cox} I.,  {Torres-Alba} N.,  {Marchesi} S.,  {Zhao} X.,  {Ajello} M.,
  {Pizzetti} A.,   {Silver} R.,  2023, \mn@doi [arXiv e-prints]
  {10.48550/arXiv.2301.07142}, \href
  {https://ui.adsabs.harvard.edu/abs/2023arXiv230107142C} {p. arXiv:2301.07142}

\bibitem[\protect\citeauthoryear{{Di Gesu} et~al.,}{{Di Gesu}
  et~al.}{2015}]{DiGesu_2015}
{Di Gesu} L.,  et~al., 2015, \mn@doi [\aap] {10.1051/0004-6361/201525934},
  \href {https://ui.adsabs.harvard.edu/abs/2015A&A...579A..42D} {579, A42}

\bibitem[\protect\citeauthoryear{{Ebrero}, {Dom{\v{c}}ek}, {Kriss}  \&
  {Kaastra}}{{Ebrero} et~al.}{2021}]{Ebrero_2021}
{Ebrero} J.,  {Dom{\v{c}}ek} V.,  {Kriss} G.~A.,   {Kaastra} J.~S.,  2021,
  \mn@doi [\aap] {10.1051/0004-6361/202040045}, \href
  {https://ui.adsabs.harvard.edu/abs/2021A&A...653A.125E} {653, A125}

\bibitem[\protect\citeauthoryear{{Elvis}, {Risaliti}, {Nicastro}, {Miller},
  {Fiore}  \& {Puccetti}}{{Elvis} et~al.}{2004}]{Elvis_2004}
{Elvis} M.,  {Risaliti} G.,  {Nicastro} F.,  {Miller} J.~M.,  {Fiore} F.,
  {Puccetti} S.,  2004, \mn@doi [\apjl] {10.1086/424380}, \href
  {https://ui.adsabs.harvard.edu/abs/2004ApJ...615L..25E} {615, L25}

\bibitem[\protect\citeauthoryear{{Gallo} et~al.,}{{Gallo}
  et~al.}{2015}]{Gallo_2015}
{Gallo} L.~C.,  et~al., 2015, \mn@doi [\mnras] {10.1093/mnras/stu2108}, \href
  {https://ui.adsabs.harvard.edu/abs/2015MNRAS.446..633G} {446, 633}

\bibitem[\protect\citeauthoryear{{Gallo}, {Gonzalez}  \& {Miller}}{{Gallo}
  et~al.}{2021}]{Gallo_2021}
{Gallo} L.~C.,  {Gonzalez} A.~G.,   {Miller} J.~M.,  2021, \mn@doi [\apjl]
  {10.3847/2041-8213/abdcb5}, \href
  {https://ui.adsabs.harvard.edu/abs/2021ApJ...908L..33G} {908, L33}

\bibitem[\protect\citeauthoryear{{Gallo}, {Miller}  \& {Costantini}}{{Gallo}
  et~al.}{2023}]{Gallo_2023}
{Gallo} L.~C.,  {Miller} J.~M.,   {Costantini} E.,  2023, \mn@doi [arXiv
  e-prints] {10.48550/arXiv.2302.10930}, \href
  {https://ui.adsabs.harvard.edu/abs/2023arXiv230210930G} {p. arXiv:2302.10930}

\bibitem[\protect\citeauthoryear{{Haardt} \& {Maraschi}}{{Haardt} \&
  {Maraschi}}{1991}]{Haardt_1991}
{Haardt} F.,  {Maraschi} L.,  1991, \mn@doi [\apjl] {10.1086/186171}, \href
  {https://ui.adsabs.harvard.edu/abs/1991ApJ...380L..51H} {380, L51}

\bibitem[\protect\citeauthoryear{{Haardt} \& {Maraschi}}{{Haardt} \&
  {Maraschi}}{1993}]{Haardt_1993}
{Haardt} F.,  {Maraschi} L.,  1993, \mn@doi [\apj] {10.1086/173020}, \href
  {https://ui.adsabs.harvard.edu/abs/1993ApJ...413..507H} {413, 507}

\bibitem[\protect\citeauthoryear{{Hancock}, {Young}  \& {Chainakun}}{{Hancock}
  et~al.}{2023}]{Hancock_2023}
{Hancock} S.,  {Young} A.~J.,   {Chainakun} P.,  2023, \mn@doi [\mnras]
  {10.1093/mnras/stad144}, \href
  {https://ui.adsabs.harvard.edu/abs/2023MNRAS.520..180H} {520, 180}

\bibitem[\protect\citeauthoryear{{Kaastra}}{{Kaastra}}{2017}]{Kaastra_2017}
{Kaastra} J.~S.,  2017, \mn@doi [\aap] {10.1051/0004-6361/201629319}, \href
  {https://ui.adsabs.harvard.edu/abs/2017A&A...605A..51K} {605, A51}

\bibitem[\protect\citeauthoryear{{Kaastra}, {Mewe}  \&
  {Nieuwenhuijzen}}{{Kaastra} et~al.}{1996}]{Kaastra_1996}
{Kaastra} J.~S.,  {Mewe} R.,   {Nieuwenhuijzen} H.,  1996, in UV and X-ray
  Spectroscopy of Astrophysical and Laboratory Plasmas. pp 411--414

\bibitem[\protect\citeauthoryear{{Kaastra} et~al.,}{{Kaastra}
  et~al.}{2012}]{Kaastra_2012}
{Kaastra} J.~S.,  et~al., 2012, \mn@doi [\aap] {10.1051/0004-6361/201118161},
  \href {https://ui.adsabs.harvard.edu/abs/2012A&A...539A.117K} {539, A117}

\bibitem[\protect\citeauthoryear{{Kaastra} et~al.,}{{Kaastra}
  et~al.}{2014}]{Kaastra_2014}
{Kaastra} J.~S.,  et~al., 2014, \mn@doi [Science] {10.1126/science.1253787},
  \href {https://ui.adsabs.harvard.edu/abs/2014Sci...345...64K} {345, 64}

\bibitem[\protect\citeauthoryear{{Kaastra}, {Raassen}, {de Plaa}  \&
  {Gu}}{{Kaastra} et~al.}{2020}]{Kaastra_2020}
{Kaastra} J.~S.,  {Raassen} A.~J.~J.,  {de Plaa} J.,   {Gu} L.,  2020, {SPEX
  X-ray spectral fitting package}, Zenodo, \mn@doi{10.5281/zenodo.4384188}

\bibitem[\protect\citeauthoryear{{Kang} \& {Wang}}{{Kang} \&
  {Wang}}{2022}]{Kang_2022}
{Kang} J.-L.,  {Wang} J.-X.,  2022, \mn@doi [\apj] {10.3847/1538-4357/ac5d49},
  \href {https://ui.adsabs.harvard.edu/abs/2022ApJ...929..141K} {929, 141}

\bibitem[\protect\citeauthoryear{{Kang} \& {Wang}}{{Kang} \&
  {Wang}}{2023}]{Kang2023}
{Kang} J.-L.,  {Wang} J.-X.,  2023, \mn@doi [\mnras] {10.1093/mnras/stac3598},
  \href {https://ui.adsabs.harvard.edu/abs/2023MNRAS.519.3635K} {519, 3635}

\bibitem[\protect\citeauthoryear{{Kang}, {Wang}  \& {Kang}}{{Kang}
  et~al.}{2021}]{Kang_2021}
{Kang} J.-L.,  {Wang} J.-X.,   {Kang} W.-Y.,  2021, \mn@doi [\mnras]
  {10.1093/mnras/stab039}, \href
  {https://ui.adsabs.harvard.edu/abs/2021MNRAS.502...80K} {502, 80}

\bibitem[\protect\citeauthoryear{{Krongold}, {Nicastro}, {Elvis}, {Brickhouse},
  {Binette}, {Mathur}  \& {Jim{\'e}nez-Bail{\'o}n}}{{Krongold}
  et~al.}{2007}]{Krongold_2007}
{Krongold} Y.,  {Nicastro} F.,  {Elvis} M.,  {Brickhouse} N.,  {Binette} L.,
  {Mathur} S.,   {Jim{\'e}nez-Bail{\'o}n} E.,  2007, \mn@doi [\apj]
  {10.1086/512476}, \href
  {https://ui.adsabs.harvard.edu/abs/2007ApJ...659.1022K} {659, 1022}

\bibitem[\protect\citeauthoryear{{Krongold} et~al.,}{{Krongold}
  et~al.}{2010}]{Krongold_2010}
{Krongold} Y.,  et~al., 2010, \mn@doi [\apj] {10.1088/0004-637X/710/1/360},
  \href {https://ui.adsabs.harvard.edu/abs/2010ApJ...710..360K} {710, 360}

\bibitem[\protect\citeauthoryear{{Laha}, {Guainazzi}, {Dewangan}, {Chakravorty}
   \& {Kembhavi}}{{Laha} et~al.}{2014}]{Laha_2014}
{Laha} S.,  {Guainazzi} M.,  {Dewangan} G.~C.,  {Chakravorty} S.,   {Kembhavi}
  A.~K.,  2014, \mn@doi [\mnras] {10.1093/mnras/stu669}, \href
  {https://ui.adsabs.harvard.edu/abs/2014MNRAS.441.2613L} {441, 2613}

\bibitem[\protect\citeauthoryear{{Lamer}, {Uttley}  \& {McHardy}}{{Lamer}
  et~al.}{2003}]{Lamer_2003}
{Lamer} G.,  {Uttley} P.,   {McHardy} I.~M.,  2003, \mn@doi [\mnras]
  {10.1046/j.1365-8711.2003.06759.x}, \href
  {https://ui.adsabs.harvard.edu/abs/2003MNRAS.342L..41L} {342, L41}

\bibitem[\protect\citeauthoryear{{Lobban}, {Turner}, {Reeves}, {Braito}  \&
  {Miller}}{{Lobban} et~al.}{2020}]{Lobban_2020}
{Lobban} A.~P.,  {Turner} T.~J.,  {Reeves} J.~N.,  {Braito} V.,   {Miller} L.,
  2020, \mn@doi [\mnras] {10.1093/mnras/staa1008}, \href
  {https://ui.adsabs.harvard.edu/abs/2020MNRAS.494.5056L} {494, 5056}

\bibitem[\protect\citeauthoryear{{Longinotti} et~al.,}{{Longinotti}
  et~al.}{2013}]{Longinotti_2013}
{Longinotti} A.~L.,  et~al., 2013, \mn@doi [\apj]
  {10.1088/0004-637X/766/2/104}, \href
  {https://ui.adsabs.harvard.edu/abs/2013ApJ...766..104L} {766, 104}

\bibitem[\protect\citeauthoryear{{Maiolino} et~al.,}{{Maiolino}
  et~al.}{2010}]{Maiolino_2010}
{Maiolino} R.,  et~al., 2010, \mn@doi [\aap] {10.1051/0004-6361/200913985},
  \href {https://ui.adsabs.harvard.edu/abs/2010A&A...517A..47M} {517, A47}

\bibitem[\protect\citeauthoryear{{Mao} et~al.,}{{Mao} et~al.}{2019}]{Mao_2019}
{Mao} J.,  et~al., 2019, \mn@doi [\aap] {10.1051/0004-6361/201833191}, \href
  {https://ui.adsabs.harvard.edu/abs/2019A&A...621A..99M} {621, A99}

\bibitem[\protect\citeauthoryear{{Markowitz} \& {Edelson}}{{Markowitz} \&
  {Edelson}}{2004}]{Markowitz_2004}
{Markowitz} A.,  {Edelson} R.,  2004, \mn@doi [\apj] {10.1086/425559}, \href
  {https://ui.adsabs.harvard.edu/abs/2004ApJ...617..939M} {617, 939}

\bibitem[\protect\citeauthoryear{{Markowitz}, {Krumpe}  \&
  {Nikutta}}{{Markowitz} et~al.}{2014}]{Markowitz_2014}
{Markowitz} A.~G.,  {Krumpe} M.,   {Nikutta} R.,  2014, \mn@doi [\mnras]
  {10.1093/mnras/stt2492}, \href
  {https://ui.adsabs.harvard.edu/abs/2014MNRAS.439.1403M} {439, 1403}

\bibitem[\protect\citeauthoryear{{Mehdipour} et~al.,}{{Mehdipour}
  et~al.}{2017}]{Mehdipour_2017}
{Mehdipour} M.,  et~al., 2017, \mn@doi [\aap] {10.1051/0004-6361/201731175},
  \href {https://ui.adsabs.harvard.edu/abs/2017A&A...607A..28M} {607, A28}

\bibitem[\protect\citeauthoryear{{Meyer} et~al.,}{{Meyer}
  et~al.}{2004}]{Meyer_2004}
{Meyer} M.~J.,  et~al., 2004, \mn@doi [\mnras]
  {10.1111/j.1365-2966.2004.07710.x}, \href
  {https://ui.adsabs.harvard.edu/abs/2004MNRAS.350.1195M} {350, 1195}

\bibitem[\protect\citeauthoryear{{Miniutti} et~al.,}{{Miniutti}
  et~al.}{2014}]{Miniutti_2014}
{Miniutti} G.,  et~al., 2014, \mn@doi [\mnras] {10.1093/mnras/stt2005}, \href
  {https://ui.adsabs.harvard.edu/abs/2014MNRAS.437.1776M} {437, 1776}

\bibitem[\protect\citeauthoryear{{Parker} et~al.,}{{Parker}
  et~al.}{2014}]{Parker_2014}
{Parker} M.~L.,  et~al., 2014, \mn@doi [\mnras] {10.1093/mnras/stu1246}, \href
  {https://ui.adsabs.harvard.edu/abs/2014MNRAS.443.1723P} {443, 1723}

\bibitem[\protect\citeauthoryear{{Parker} et~al.,}{{Parker}
  et~al.}{2019}]{Parker_2019}
{Parker} M.~L.,  et~al., 2019, \mn@doi [\mnras] {10.1093/mnras/stz2566}, \href
  {https://ui.adsabs.harvard.edu/abs/2019MNRAS.490..683P} {490, 683}

\bibitem[\protect\citeauthoryear{{Puccetti}, {Fiore}, {Risaliti}, {Capalbi},
  {Elvis}  \& {Nicastro}}{{Puccetti} et~al.}{2007}]{Puccetti_2007}
{Puccetti} S.,  {Fiore} F.,  {Risaliti} G.,  {Capalbi} M.,  {Elvis} M.,
  {Nicastro} F.,  2007, \mn@doi [\mnras] {10.1111/j.1365-2966.2007.11634.x},
  \href {https://ui.adsabs.harvard.edu/abs/2007MNRAS.377..607P} {377, 607}

\bibitem[\protect\citeauthoryear{{Reeves}, {Lobban}  \& {Pounds}}{{Reeves}
  et~al.}{2018}]{Reeves_2018}
{Reeves} J.~N.,  {Lobban} A.,   {Pounds} K.~A.,  2018, \mn@doi [\apj]
  {10.3847/1538-4357/aaa776}, \href
  {https://ui.adsabs.harvard.edu/abs/2018ApJ...854...28R} {854, 28}

\bibitem[\protect\citeauthoryear{{Risaliti}, {Elvis}, {Fabbiano}, {Baldi}  \&
  {Zezas}}{{Risaliti} et~al.}{2005}]{Risaliti_2005}
{Risaliti} G.,  {Elvis} M.,  {Fabbiano} G.,  {Baldi} A.,   {Zezas} A.,  2005,
  \mn@doi [\apjl] {10.1086/430252}, \href
  {https://ui.adsabs.harvard.edu/abs/2005ApJ...623L..93R} {623, L93}

\bibitem[\protect\citeauthoryear{{Risaliti}, {Elvis}, {Fabbiano}, {Baldi},
  {Zezas}  \& {Salvati}}{{Risaliti} et~al.}{2007}]{Risaliti_2007}
{Risaliti} G.,  {Elvis} M.,  {Fabbiano} G.,  {Baldi} A.,  {Zezas} A.,
  {Salvati} M.,  2007, \mn@doi [\apjl] {10.1086/517884}, \href
  {https://ui.adsabs.harvard.edu/abs/2007ApJ...659L.111R} {659, L111}

\bibitem[\protect\citeauthoryear{{Risaliti}, {Young}  \& {Elvis}}{{Risaliti}
  et~al.}{2009}]{Risaliti_2009}
{Risaliti} G.,  {Young} M.,   {Elvis} M.,  2009, \mn@doi [\apjl]
  {10.1088/0004-637X/700/1/L6}, \href
  {https://ui.adsabs.harvard.edu/abs/2009ApJ...700L...6R} {700, L6}

\bibitem[\protect\citeauthoryear{{Risaliti}, {Nardini}, {Salvati}, {Elvis},
  {Fabbiano}, {Maiolino}, {Pietrini}  \& {Torricelli-Ciamponi}}{{Risaliti}
  et~al.}{2011}]{Risaliti_2011}
{Risaliti} G.,  {Nardini} E.,  {Salvati} M.,  {Elvis} M.,  {Fabbiano} G.,
  {Maiolino} R.,  {Pietrini} P.,   {Torricelli-Ciamponi} G.,  2011, \mn@doi
  [\mnras] {10.1111/j.1365-2966.2010.17503.x}, \href
  {https://ui.adsabs.harvard.edu/abs/2011MNRAS.410.1027R} {410, 1027}

\bibitem[\protect\citeauthoryear{{Rivers}, {Markowitz}  \&
  {Rothschild}}{{Rivers} et~al.}{2011}]{Rivers_2011}
{Rivers} E.,  {Markowitz} A.,   {Rothschild} R.,  2011, \mn@doi [\apjl]
  {10.1088/2041-8205/742/2/L29}, \href
  {https://ui.adsabs.harvard.edu/abs/2011ApJ...742L..29R} {742, L29}

\bibitem[\protect\citeauthoryear{{Sako} et~al.,}{{Sako}
  et~al.}{2001}]{Sako_2001}
{Sako} M.,  et~al., 2001, \mn@doi [\aap] {10.1051/0004-6361:20000081}, \href
  {https://ui.adsabs.harvard.edu/abs/2001A&A...365L.168S} {365, L168}

\bibitem[\protect\citeauthoryear{{Sanfrutos}, {Miniutti},
  {Ag{\'\i}s-Gonz{\'a}lez}, {Fabian}, {Miller}, {Panessa}  \&
  {Zoghbi}}{{Sanfrutos} et~al.}{2013}]{Sanfrutos_2013}
{Sanfrutos} M.,  {Miniutti} G.,  {Ag{\'\i}s-Gonz{\'a}lez} B.,  {Fabian} A.~C.,
  {Miller} J.~M.,  {Panessa} F.,   {Zoghbi} A.,  2013, \mn@doi [\mnras]
  {10.1093/mnras/stt1675}, \href
  {https://ui.adsabs.harvard.edu/abs/2013MNRAS.436.1588S} {436, 1588}

\bibitem[\protect\citeauthoryear{{Sarma}, {Tripathi}, {Misra}, {Dewangan},
  {Pathak}  \& {Sarma}}{{Sarma} et~al.}{2015}]{Sarma2015}
{Sarma} R.,  {Tripathi} S.,  {Misra} R.,  {Dewangan} G.,  {Pathak} A.,
  {Sarma} J.~K.,  2015, \mn@doi [\mnras] {10.1093/mnras/stv005}, \href
  {https://ui.adsabs.harvard.edu/abs/2015MNRAS.448.1541S} {448, 1541}

\bibitem[\protect\citeauthoryear{{Silva}, {Uttley}  \& {Costantini}}{{Silva}
  et~al.}{2016}]{Silva_2016}
{Silva} C.~V.,  {Uttley} P.,   {Costantini} E.,  2016, \mn@doi [\aap]
  {10.1051/0004-6361/201628555}, \href
  {https://ui.adsabs.harvard.edu/abs/2016A&A...596A..79S} {596, A79}

\bibitem[\protect\citeauthoryear{{Sobolewska} \& {Papadakis}}{{Sobolewska} \&
  {Papadakis}}{2009}]{Sobolewska_2009}
{Sobolewska} M.~A.,  {Papadakis} I.~E.,  2009, \mn@doi [\mnras]
  {10.1111/j.1365-2966.2009.15382.x}, \href
  {https://ui.adsabs.harvard.edu/abs/2009MNRAS.399.1597S} {399, 1597}

\bibitem[\protect\citeauthoryear{{Soldi} et~al.,}{{Soldi}
  et~al.}{2014}]{Soldi_2014}
{Soldi} S.,  et~al., 2014, \mn@doi [\aap] {10.1051/0004-6361/201322653}, \href
  {https://ui.adsabs.harvard.edu/abs/2014A&A...563A..57S} {563, A57}

\bibitem[\protect\citeauthoryear{{Steenbrugge}, {Kaastra}, {de Vries}  \&
  {Edelson}}{{Steenbrugge} et~al.}{2003}]{Steenbrugge2003}
{Steenbrugge} K.~C.,  {Kaastra} J.~S.,  {de Vries} C.~P.,   {Edelson} R.,
  2003, \mn@doi [\aap] {10.1051/0004-6361:20030261}, \href
  {https://ui.adsabs.harvard.edu/abs/2003A&A...402..477S} {402, 477}

\bibitem[\protect\citeauthoryear{{Steenbrugge}, {Fenov{\v{c}}{\'\i}k},
  {Kaastra}, {Costantini}  \& {Verbunt}}{{Steenbrugge}
  et~al.}{2009}]{Steenbrugge_2009}
{Steenbrugge} K.~C.,  {Fenov{\v{c}}{\'\i}k} M.,  {Kaastra} J.~S.,  {Costantini}
  E.,   {Verbunt} F.,  2009, \mn@doi [\aap] {10.1051/0004-6361/200810416},
  \href {https://ui.adsabs.harvard.edu/abs/2009A&A...496..107S} {496, 107}

\bibitem[\protect\citeauthoryear{{Str{\"u}der} et~al.,}{{Str{\"u}der}
  et~al.}{2001}]{Struder_2001_PN}
{Str{\"u}der} L.,  et~al., 2001, \mn@doi [\aap] {10.1051/0004-6361:20000066},
  \href {https://ui.adsabs.harvard.edu/abs/2001A&A...365L..18S} {365, L18}

\bibitem[\protect\citeauthoryear{{Turner}, {Reeves}, {Kraemer}  \&
  {Miller}}{{Turner} et~al.}{2008}]{Turner_2008}
{Turner} T.~J.,  {Reeves} J.~N.,  {Kraemer} S.~B.,   {Miller} L.,  2008,
  \mn@doi [\aap] {10.1051/0004-6361:20078808}, \href
  {https://ui.adsabs.harvard.edu/abs/2008A&A...483..161T} {483, 161}

\bibitem[\protect\citeauthoryear{{Turner}, {Miller}, {Kraemer}  \&
  {Reeves}}{{Turner} et~al.}{2011}]{Turner_2011}
{Turner} T.~J.,  {Miller} L.,  {Kraemer} S.~B.,   {Reeves} J.~N.,  2011,
  \mn@doi [\apj] {10.1088/0004-637X/733/1/48}, \href
  {https://ui.adsabs.harvard.edu/abs/2011ApJ...733...48T} {733, 48}

\bibitem[\protect\citeauthoryear{{Turner}, {Reeves}, {Braito}, {Lobban},
  {Kraemer}  \& {Miller}}{{Turner} et~al.}{2018}]{Turner_2018}
{Turner} T.~J.,  {Reeves} J.~N.,  {Braito} V.,  {Lobban} A.,  {Kraemer} S.,
  {Miller} L.,  2018, \mn@doi [\mnras] {10.1093/mnras/sty2447}, \href
  {https://ui.adsabs.harvard.edu/abs/2018MNRAS.481.2470T} {481, 2470}

\bibitem[\protect\citeauthoryear{{Wang} et~al.,}{{Wang}
  et~al.}{2022}]{Wang_2022}
{Wang} Y.,  et~al., 2022, \mn@doi [\aap] {10.1051/0004-6361/202141599}, \href
  {https://ui.adsabs.harvard.edu/abs/2022A&A...657A..77W} {657, A77}

\bibitem[\protect\citeauthoryear{{Wilkins} \& {Gallo}}{{Wilkins} \&
  {Gallo}}{2015}]{Wilkins_2015}
{Wilkins} D.~R.,  {Gallo} L.~C.,  2015, \mn@doi [\mnras]
  {10.1093/mnras/stv162}, \href
  {https://ui.adsabs.harvard.edu/abs/2015MNRAS.449..129W} {449, 129}

\bibitem[\protect\citeauthoryear{{Willingale}, {Starling}, {Beardmore},
  {Tanvir}  \& {O'Brien}}{{Willingale} et~al.}{2013}]{Willingale_2013}
{Willingale} R.,  {Starling} R.~L.~C.,  {Beardmore} A.~P.,  {Tanvir} N.~R.,
  {O'Brien} P.~T.,  2013, \mn@doi [\mnras] {10.1093/mnras/stt175}, \href
  {https://ui.adsabs.harvard.edu/abs/2013MNRAS.431..394W} {431, 394}

\bibitem[\protect\citeauthoryear{{Wu}, {Wang}, {Cai}, {Kang}, {Liu}  \&
  {Cai}}{{Wu} et~al.}{2020}]{Wu_2020}
{Wu} Y.-J.,  {Wang} J.-X.,  {Cai} Z.-Y.,  {Kang} J.-L.,  {Liu} T.,   {Cai} Z.,
  2020, \mn@doi [Science China Physics, Mechanics, and Astronomy]
  {10.1007/s11433-020-1611-7}, \href
  {https://ui.adsabs.harvard.edu/abs/2020SCPMA..6329512W} {63, 129512}

\bibitem[\protect\citeauthoryear{{de Plaa}, {Kaastra}, {Tamura},
  {Pointecouteau}, {Mendez}  \& {Peterson}}{{de Plaa}
  et~al.}{2004}]{dePlaa_2004}
{de Plaa} J.,  {Kaastra} J.~S.,  {Tamura} T.,  {Pointecouteau} E.,  {Mendez}
  M.,   {Peterson} J.~R.,  2004, \mn@doi [\aap]
  {10.1051/0004-6361:2004717010.48550/arXiv.astro-ph/0405307}, \href
  {https://ui.adsabs.harvard.edu/abs/2004A&A...423...49D} {423, 49}

\bibitem[\protect\citeauthoryear{{den Herder} et~al.,}{{den Herder}
  et~al.}{2001}]{Herder_2001_RGS}
{den Herder} J.~W.,  et~al., 2001, \mn@doi [\aap] {10.1051/0004-6361:20000058},
  \href {https://ui.adsabs.harvard.edu/abs/2001A&A...365L...7D} {365, L7}

\makeatother
\end{thebibliography}

\bsp	
\label{lastpage}
\end{document}